%% file: ms_final.tex
\newcommand{\cha}{{\it Chandra}\,}

\documentclass[apj]{emulateapj}

\received{2006 October 12}
\accepted{2007 January 22}

\shortauthors{Maggio et al.}
\shorttitle{Coronal abundances in Orion}

\begin{document}

\title{Coronal Abundances in Orion Nebula Cluster Stars}

\author{A. Maggio\altaffilmark{1}, E. Flaccomio\altaffilmark{1},
F. Favata\altaffilmark{2}, G. Micela\altaffilmark{1}, S.
Sciortino\altaffilmark{1}, \\E. D. Feigelson\altaffilmark{3}, K.
V. Getman\altaffilmark{3} }

\altaffiltext{1}{INAF -- Osservatorio Astronomico di Palermo
Giuseppe S. Vaiana, Piazza del Parlamento 1, I--90134 Palermo,
Italy; maggio@astropa.unipa.it, ettorefo@astropa.unipa.it,
micela@astropa.unipa.it, sciorti@astropa.unipa.it.}
\altaffiltext{2}{Astrophysics Division - Research and Space Science
Department of ESA, ESTEC, Noordwijk, NL} \altaffiltext{3}{Dept. of
Astronomy \& Astrophysics, Pennsylvania State University, University
Park PA 16802, USA}

\begin{abstract}

Following the \cha~Orion Ultradeep Project (COUP) observation, we
have studied the chemical composition of the hot plasma in a sample
of 146 X-ray bright pre-main sequence stars in the Orion Nebula
Cluster (ONC). We report measurements of individual element
abundances for a subsample of 86
slightly-absorbed and bright X-ray sources, using low resolution
X-ray spectra obtained from the Chandra ACIS instrument. The X-ray
emission originates from a plasma with temperatures and elemental
abundances very similar to those of active coronae
in older stars.
A clear pattern of abundances vs.\ First Ionization
Potential (FIP) is evident
if solar photospheric abundances are assumed as reference. 
The results are validated by extensive
simulations.   The observed abundance distributions are compatible
with a single pattern of abundances for all stars, although a weak
dependence on flare loop size may be present.  The abundance of
calcium is the only one which appears to vary 
substantially between stars, but this quantity is affected by
relatively large uncertainties.

The ensemble properties of the X-ray bright COUP sources confirm
that the iron in the emitting plasma is underabundant with respect
to both the solar composition and to the average stellar
photospheric values. Comparison of the present plasma abundances
with those of the stellar photospheres and those of the gaseous
component of the nebula, indicates a good agreement for all the
other elements with available measurements, and in particular for
the high-FIP elements (Ne, Ar, O, and S) and for the low-FIP element
Si. We conclude that there is evidence of a significant chemical
fractionation effect only for iron, which appears to be depleted 
by a factor 1.5--3 with respect to the stellar composition.

\end{abstract}


\section{Introduction}
\label{sec:intro}

Chemical composition is one of the key properties of astrophysical
environments, fundamental for classifying stellar populations and
studying the evolutionary history of galactic chemistry over
different spatial scales. Several processes altering element
abundances in the vicinity of individual stars are particularly
efficient in the early phases of stellar evolution: selective
trapping in grains, high-energy photon and particle irradiation of
the circumstellar medium, mass exchange between stars and
protoplanetary disks via accretion and outflows, and fractionation
effects in stellar coronae and magnetospheres. In pre-main sequence
systems, these physical processes occur on short time scales and
cause dramatic changes as new planetary systems form out of the
circumstellar nebula. One of the widely discussed issues is why
planet-hosting stars appear to be characterized by a metallicity
higher (by 0.24 dex, on average) than stars without planets
(\citealp{sim+05}, and references therein).

The Orion Nebula Cluster (ONC) is one of the best studied star
forming regions in the sky, and the chemical composition of the
associated \ion{H}{2} region has been historically considered a
standard reference for ionized gas in the nearby Galaxy \citep{e+05}.
Optical spectroscopy of the nebula is the traditional
approach employed to study the abundances of important elements in
this environment, as well as in other star forming regions, because
of the difficulty in obtaining photospheric abundance measurements
for faint, rapidly rotating young stars. Abundance studies of Orion
stars have been performed mainly on B main-sequence members 
\citep*{cl92,cl94,sd+06,chl06}.
Only a few measurements are available for
slowly-rotating F and G stars \citep*{csl98} and
K-M members \citep{cs05}.

An alternative way to determine the chemical composition of
late-type stars is emerging from X-ray spectroscopy 
(\citealp{g04}, Favata \& Micela 2003, and references therein). 
The ONC was selected in 2003 for a very long observation program in X-rays 
with the \cha~satellite, known as the \cha~Orion Ultradeep Project (COUP,
\citealp{g+05}). This program is providing an unprecedented
wealth of information about the stellar population of the ONC and
various characteristics of this prototypical stellar and planetary
nursery.

Among the salient global properties of the ONC, \citet{fgt+05}
noted the presence of a strong spectral feature around 1\,keV
in the cumulative spectrum of all detected X-ray sources, identified
with the emission line complex due to H-like and He-like Ne ions in
hot plasma associated with $\sim 1400$ ONC members. The prominence
of this feature suggests a high abundance of Ne in the X-ray
emitting plasma, a characteristic already observed in other
magnetically active stars (see review by \citealp{g04}) together with
an apparent depletion of iron in corona with respect to the {\em
expected} photospheric composition.

This behavior is linked to the First Ionization Potential (FIP) of
the elements and possibly to the stellar activity level:
the composition of the coronal plasma in the Sun,
and particularly in long-lived magnetic structures, appears enriched by
low-FIP elements ($FIP < 10$\,eV) by about a factor 4 with respect
to photospheric values.  This is known as the FIP effect (e.g.
\citealp{fl00}), a characteristic observed also in other
low-activity stars \citep[and references therein]{fm03}. 
High-activity RS CVn-type and Algol-type
binaries stars exhibit a different behavior with a tendency for
low-FIP elements such as iron to become depleted with respect to
high-FIP elements like argon and neon.  This is called ``inverse FIP
effect'' \citep{b+01}.

The above scenario requires further investigations for several
reasons.  First, photospheric abundances are usually uncertain due to severe
NLTE and rotational broadening effects on optical spectra of active stars,
and solar values are often employed as the reference for the stellar 
coronal abundances. The solar photospheric composition itself
has been recently questioned by \citet{a05}, based on
detailed 3-D modeling of the solar atmosphere.
Moreover, the photospheric Ne abundance is not directly measurable 
even in the Sun because no Ne lines occur at optical wavelengths.

Second, the driving mechanism(s) for FIP-related
fractionation in stellar upper atmospheres still escapes a clear
understanding although some models have emerged \citep*{am98,s+99,l04}.

Third, the situation is made more complex by the presence of very
high Ne abundance ratios in two classical T Tauri stars (CTTS),
TW Hya and BP Tau, where the soft X-ray emission is often attributed
to an accretion shock rather than to magnetic activity 
\citep{khs+02,ss04,srn+05}.
An alternative explanation for these high Ne abundances has been suggested based
on an origin of the plasma in gas accreted from the circumstellar
disks where refractory elements may be depleted into solids
undergoing growth into planetary bodies \citep*{dth05}.
However, X-ray variability characteristics strongly favors CTTS
X-ray emission dominated by magnetic flares rather than accretion
\citep{svff06,sfb+06}.

The COUP observation provides us with the largest homogeneous sample
ever studied to address the issue of element abundances in X-ray
emitting plasmas associated to young stars.  Here we can exploit
ensemble statistical properties to overcome uncertainties in
individual stellar measurements. The uncertainties on abundances
derived from X-ray spectra, even those obtained with the highest
spectral resolution, may be larger than formal statistical error
bars \citep{sn04,m+05}. The large COUP
stellar sample can compensate for this difficulty.

To address these issues, we present in this paper the results of a
detailed analysis of CCD-resolution X-ray spectra of ONC members to
derive the abundances of individual elements for a large number of
young stars. In \S \ref{sec:obs}, we introduce the observation and
the sample selection, \S \ref{sec:method} is devoted to the
methodology of analysis, while the results are presented in \S
\ref{sec:res} and discussed in \S\S \ref{sec:discuss}--9.

\section{Observation and sample selection}
\label{sec:obs}

The COUP data were obtained in January 2003 with the Advanced CCD
Imaging Spectrometer (ACIS; \citealp{g+03}) on-board the
\cha~X-ray Observatory \citep{w+02} by combining six
consecutive observations of the Orion Nebula Cluster (ONC) with the
same aimpoint and roll angle. The resulting data set has a total
exposure time of $\sim 838$\,ks (9.7 days) distributed over 13.2
days. The data reduction and analysis resulted in 1616 detected
X-ray sources, of which about 1408 are associated with ONC stars
\citep{g+05}. X-ray spectra and light curves for each source
were constructed from events collected in a polygonal region
centered at the source position including $\sim 90\%$ of the
encircled energy. Here we employ the spectra and the related
auxiliary response files (ARFs), created with the {\sc
acis\_extract} software, as explained in Getman et al.

The criteria for sample selection were motivated by the need to
perform a detailed analysis of the ACIS spectra, and -- more
specifically -- to determine both the plasma temperature
distribution in the corona and the abundances of a number of
individual elements. The main requirement for this kind of analysis
is a very strong signal (noise is negligible in these cases) and an
excellent knowledge of the instrument spectral response. We select
stars with at least $5000$ total (net) counts in the 0.5-8\,keV
energy band. Sources heavily affected by pile-up (flagged ``a'' in
Table\,6 of \citealp{g+05}) are excluded but the sample includes
23 stars with ``mild'' pile-up effects (``w'' flag) for which the
usual whole extraction region was used. We exclude four early-type
stars ($T_{\rm eff} > 10,000$\,K) as their X-ray emission likely
originates in wind shocks rather than in a corona. However, we
have retained 34 stars with unknown $T_{\rm eff}$. The final sample
contains 146 COUP sources listed in Table \ref{tab:opt}; about half of
them have more than $8000$ total counts, and $55$ exceed $10^4$ counts. 

Further {\it a posteriori} selection was made based on the amount of
interstellar absorption exhibited in the X-ray spectra. Reliable
determination of the abundances of some elements is possible only
for slightly absorbed sources.  We will focus our attention on the
86 sources with intervening hydrogen column densities $N_H < 6
\times 10^{21}$\,cm$^{-2}$.

\begin{figure}[!th]
\plottwo{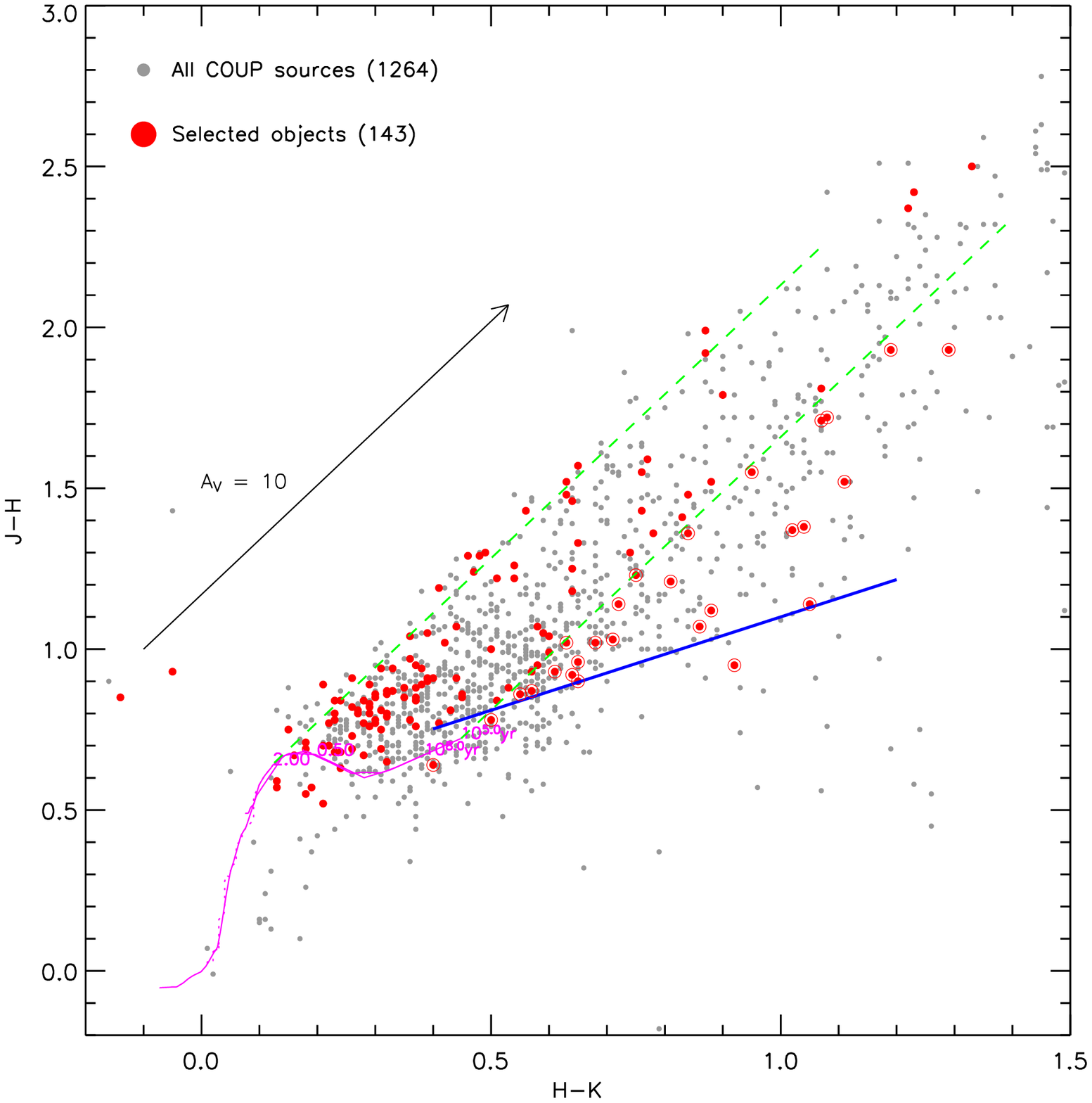}{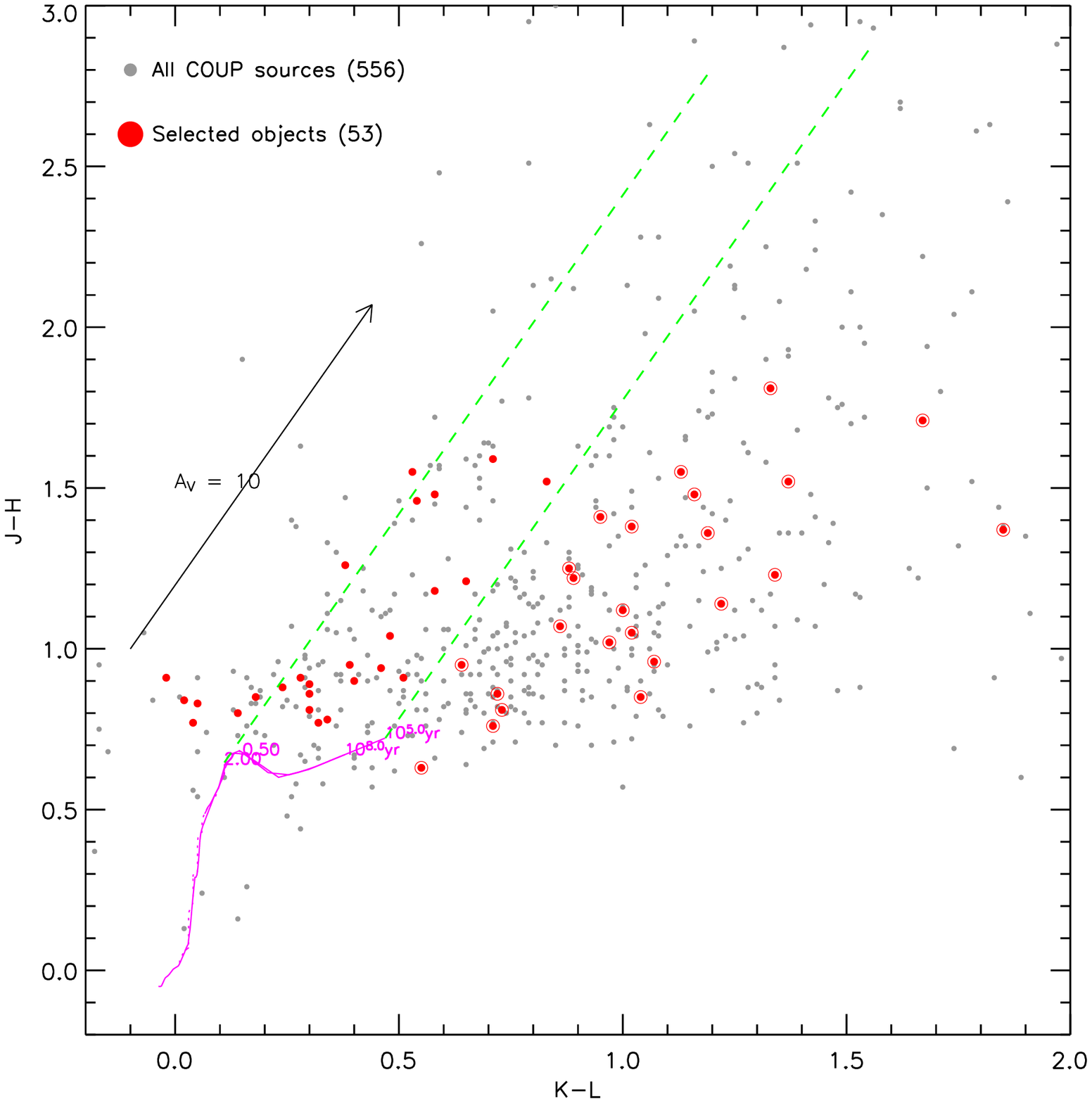}

\caption{ Near-infrared color-color diagrams of the COUP sources
with the stars in the present study indicated by filled red circles.
The number of stars in the two diagrams differ because of incomplete
IR photometric data for some of them. In magenta, the locus of stars
with ages 1--5 Myr (\citealp*{sdf00}, using $T_{\rm eff}$-color
conversions by \citealp{kh95}). The black arrow indicates a
reddening vector corresponding to $A_{\rm v} = 10$, the dashed green
lines bracket the region where stars with no NIR excess are
expected, and the blue segment marks the locus of dereddened
classical T~Tauri stars \citep{mch97}. \label{fig:nir}}
\end{figure}

The optical properties of the 146 stars in our initial sample are
reported in Table~\ref{tab:opt}, based on the tables in \citet{g+05}.
Figure~\ref{fig:nir} shows two near-infrared (NIR)
color-color diagrams for all the COUP detected sources with
available photometric measurements in the NIR bands of interest.
These diagrams show the theoretical locus of stars at the
appropriate ONC age, and loci populated by objects with different
amounts of reddening.  Twelve sources fall to the right and below
these loci in both color-color diagrams indicating they probably
have dusty circumstellar disks in addition to reddening.  Three of
these sources plus two more without NIR photometric excesses (COUP
382, 579, 597, 758, and 1409) reveal protoplanetary disks seen in
silhouette against the bright nebula (``proplyds'') in Hubble Space
Telescope imaging \citep{kfg+05}.

Table~\ref{tab:opt} also reports the equivalent width of the 8542
\AA\ \ion{Ca}{2} line from the study of \citet{h97}.
In 14 cases, this line is in emission with equivalent width larger
than 1\,\AA, suggesting that the central star is actively accreting
material from the circumstellar medium.

\section{Analysis}
\label{sec:method}

Essentially all X-ray bright COUP sources show significant
variability of their X-ray emission level \citep{g+05,ffr+05,whf+05}.
In many cases, this variability is associated with large flares which
are well-fit by a solar-type flare model where plasma is suddenly
heated by a magnetic reconnection event and cools on timescales of
hours-to-days \citep{ffr+05}. In the present work,  we perform
the X-ray spectral analysis collecting photons over the whole
observation length which includes quiescent and flaring episodes.
This may confuse interpretation of elemental abundances, as both
temperatures, and sometimes FIP effects, vary during stellar flares
\citep*[e.g.][]{agm01,oba+04,nbg06}.
Our results thus give thermal and chemical characteristics of the
emitting plasma which are spatially and temporally averaged over for
each star.  This conflation of `quiescent' and flare conditions
may not be avoidable as there may not exist any true `quiescent'
state in very active stars such as those in the ONC because, even
during periods of little variability, the X-ray emission likely arises from a
superposition of a multitude of weaker flares (\citealp{g04} and
references therein). 
This continuous flaring paradigm implies that
the abundance properties we derive for the COUP sources probably
describe time-averaged dynamical conditions rather than static
equilibrium conditions of the X-ray emitting plasma.

We assume that the observed emission can be modeled as a
collisionally-excited plasma in ionization equilibrium, and we adopt
the emissivities predicted by the Astrophysical Plasma Emission Code
(APEC V1.3.1, \citealp{sbl+01}) in the spectral fitting process.
This choice, rather than the MEKAL emissivities \citep{mkl95}
adopted in previous COUP works, is motivated by the
significantly larger number of emission lines and more updated
atomic data of APEC vs.\ MEKAL as implemented in the XSPEC spectral
analysis package. This is especially important in our work, aimed to
derive information from line complexes due to specific atomic
species of individual elements, which can be resolved at most only
marginally in the available CCD spectra. In any case, this choice
may affect the abundance measurements of some elements, but not our
global results.

The data sets and instrument spectral response for each source are
obtained from the data reduction described by \citet{g+05}.
Special features of the COUP data processing include a $0.5-8$~keV
energy range and, for these strong sources, grouping of energy
channels such that there are at least 60 net counts in each spectral
bin.

\subsection{Spectral diagnostics of element abundances}
\label{sec:spectra}

We adopt a global spectral fitting approach with multi-temperature
models where individual element abundances are free parameters, in
addition to the temperature and volume emission measure of each
component and the interstellar hydrogen column density $N_{\rm H}$
to the star. Photoelectric absorption in the interstellar medium
(ISM) is modeled with cross-sections obtained by \citet{mm83}.
The spectral resolution of ACIS CCDs is such
that a model with two or three isothermal components usually
provides a good description of the observed spectra.  This modeling
approach provides an approximate description of the continuous
distribution of temperatures undoubtedly present in the X-ray
emitting plasma, but it is certainly adequate to the amount of
information provided by instruments with resolution power $\approx
10$--30.

\begin{figure}[!th]
\plotone{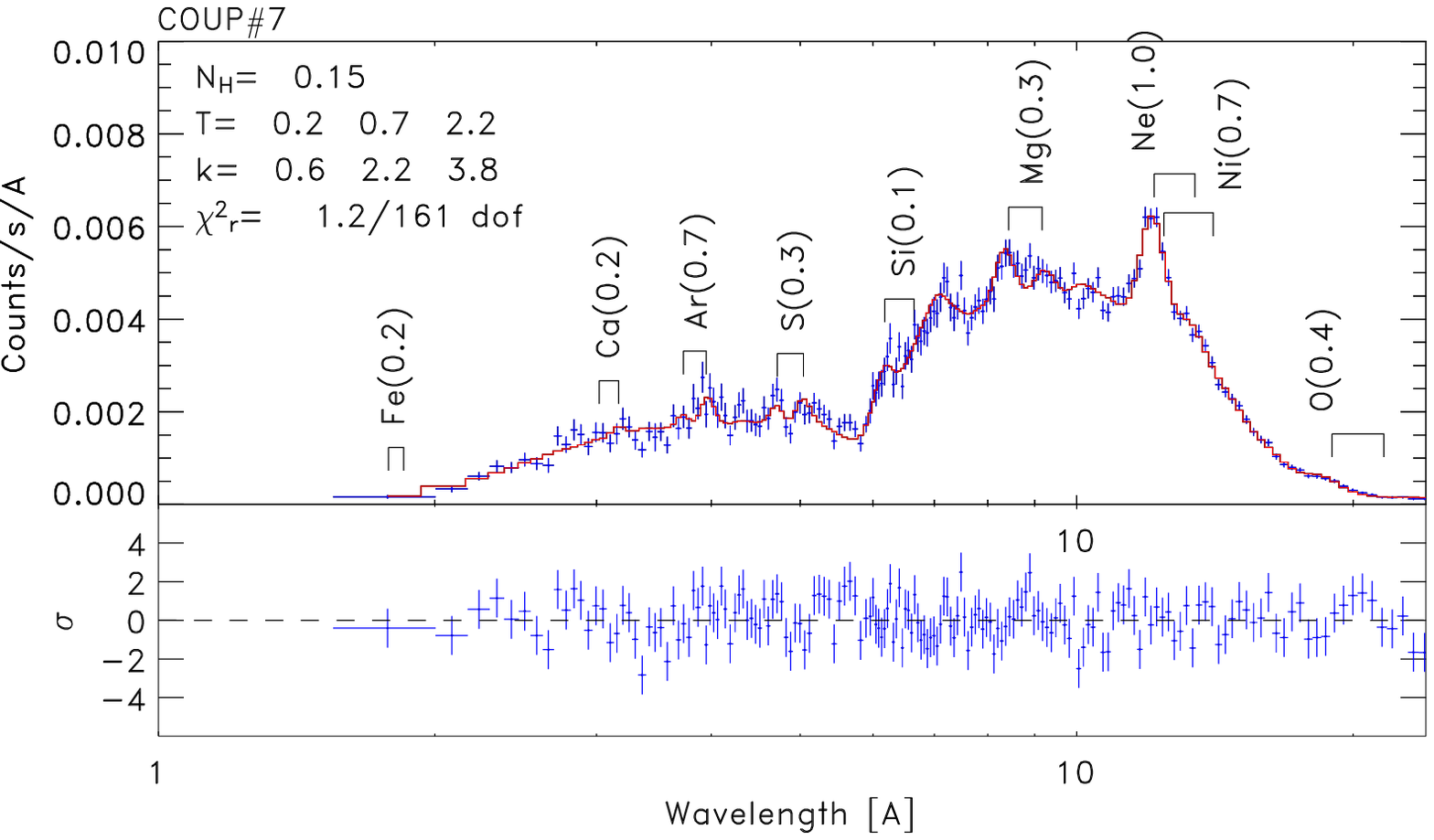}
\plotone{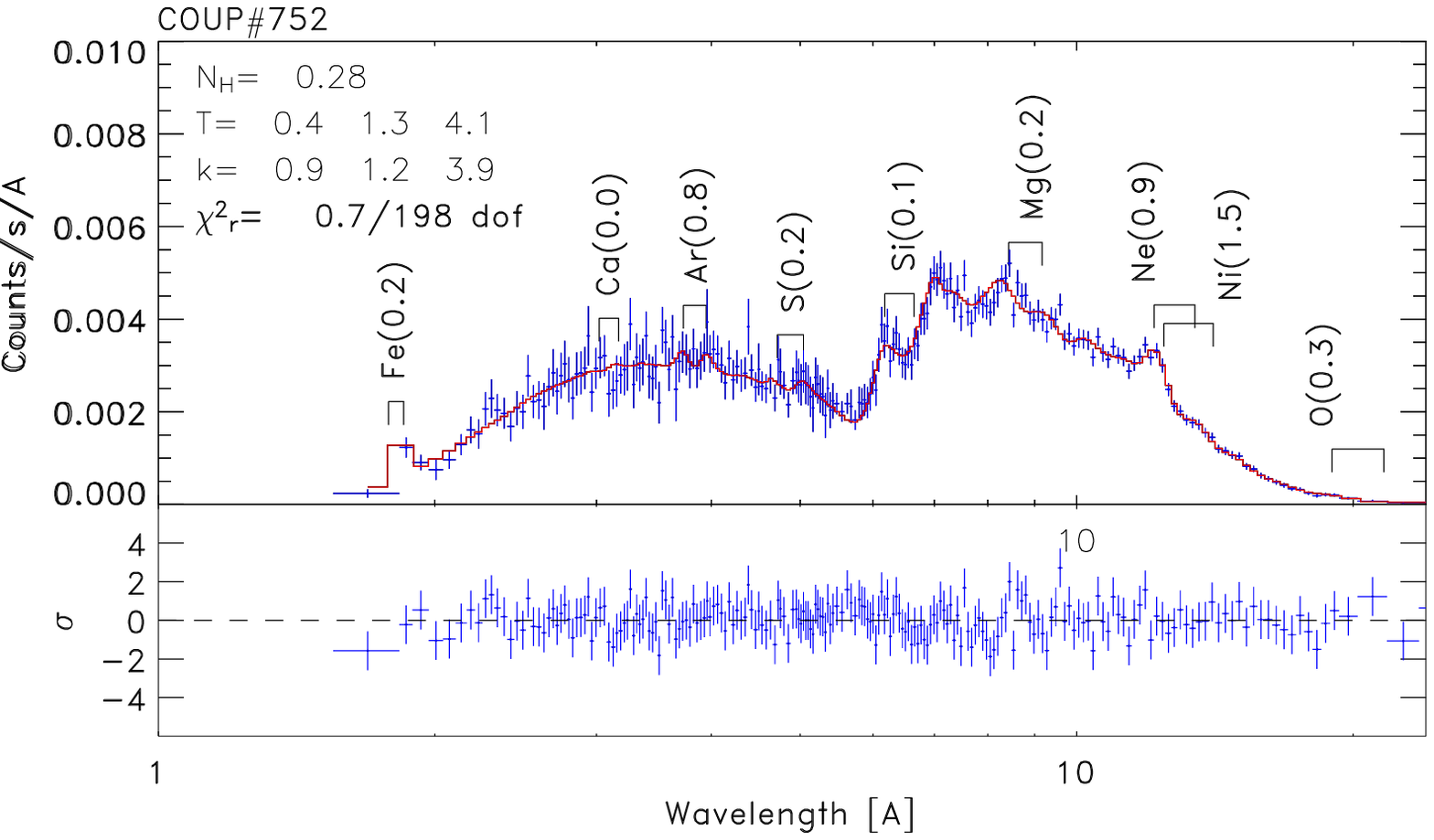}

\caption{ Examples of spectra of Orion COUP sources with major
emission line complexes labeled by their element abundances (in
solar photospheric units). The best-fit 3-T model
parameters ($N_{\rm H}$ in units of $10^{22}$\,cm$^{-2}$, $T$ in
keV, $k$ in cm$^{-5}$) and the reduced $\chi^2$ are indicated in the
top panels, while bottom panels show the residuals in units of
standard deviation. \label{fig:spsample}}
\end{figure}

For the abundance measurements, we consider the elements O, Ne, Mg,
Si, S, Ar, Ca, Fe, and Ni. At the coronal temperatures typical of
young active stars ($T \approx 10^7$\,K), all of these atoms have
important H-like or He-like ion lines in the ACIS wavelength range
(1.5--27.6\,\AA, Fig.\,\ref{fig:spsample}). Iron and Nickel are also
represented by a large number of L-shell emission lines. However, clear
spectral signatures of each element depends on several factors:
relative line emissivities, plasma emission measure $vs.$\
temperature distribution, line blending, interstellar absorption
and, of course, the abundance of each element in the plasma. Due to
the complexity of these dependencies, we have performed extensive
sets of simulations to validate our spectral fitting results.
Details on these simulations are reported in Appendix and will be
included in discussion of specific results below.

\begin{figure}[!th]
\plotone{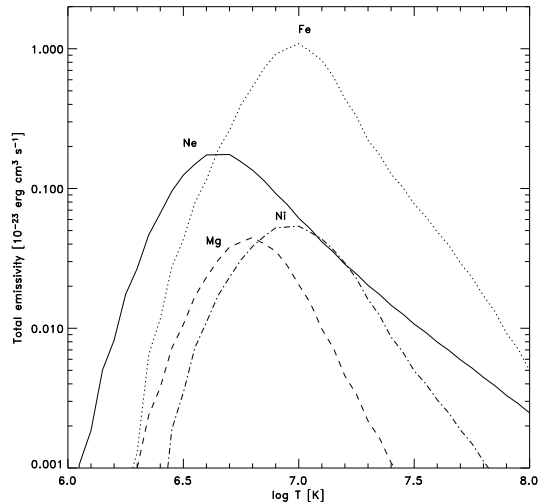}

\caption{ Total emissivity of emission lines from Ne, Mg, Fe, and Ni
ions in the wavelength range 9--14.5\AA, vs.\ temperature from the
{\it APEC} plasma model assuming the solar element abundances of
\citet{ag89}. \label{fig:emiw}}
\end{figure}

The neon abundance is especially interesting for several reasons.
First, this abundance cannot be determined in stellar photospheres
because of the lack of suitable absorption lines in optical spectra,
hence X-ray data provide the best opportunity to perform such a
measurement.  Second, neon has the highest First Ionization
Potential (FIP = 21.56\,eV) of any atoms except helium, hence its
abundance with respect to iron (with a low FIP)
provides crucial information about the stratification of elements
with different FIPs in stellar atmospheres (\S~\ref{sec:intro}).
Third, H-like and He-like Ne ions produce prominent emission lines
at energies around 1~keV where $Chandra$/ACIS sensitivity is
highest.  The correct determination of the Ne abundance is not free
of difficulties due to the proximity of its most intense emission
lines with L-shell iron and nickel lines. Figure~\ref{fig:emiw}
shows this with a plot of the integrated emissivity of Ne, Mg, Fe,
and Ni lines in the wavelength range 9--14.5\,\AA, vs.\ plasma
temperature for a solar mixture of chemical elements \citep{ag89}.
Although the Fe emissivity exceeds that of Ne for
solar abundances, when the Ne/Fe ratio is several times the solar
ratio as found in many active stars \citep{dbk+01,g04}, the strength of
the Ne lines in this spectral range becomes comparable to or greater
than that of the iron lines. Thus, Ne abundances in active stellar
coronae are sufficiently high to bring this element within the reach
of global spectral analysis, as we will illustrate shortly.

\subsection{Spectral fitting procedure}
\label{sec:good}

Our procedure starts with fitting of 2-temperature (2-T) and
3-temperature (3-T) plasma models to all the source spectra using
a $\chi^2$-minimization algorithm implemented in the {\it XSPEC} (V11.3)
package \citep{a96}.  This step is repeated a few times with
different starting values of the free parameters to avoid local
minimum $\chi^2$ solutions. An F-test is then applied to determine
whether the (usually lower) $\chi^2$ obtained with the 3-T model
represent a significantly better fit to the data, or rather the
improvement is entirely due to the larger number of free parameters
with respect to the 2-T model. We adopt 3-T results only if the
following two conditions are met: (i) the 2-T model has a poor fit
with probability $P(\chi^2) < 10$\%; and (ii) the F-test shows the
3-T model is significantly better with $P(F) < 10\%$.
The unnecessary introduction
of a third thermal component, which typically has the lowest
temperature, may alter the abundances of elements (such as O, Ne,
and Mg) with emission lines only in the low-energy tail of the
spectrum.

Our elemental abundances are scaled to the widely used solar
system abundances of \citet{ag89}. In
\S~\ref{sec:phys} we will also discuss the implications of the
recent solar composition recommended by \citet{ags05}.

This procedure resulted in 118 spectra fitted with a 2-T model and
28 spectra fitted with 3-T models.  For 18 spectra, none of the
models provides a statistically acceptable fit at the 99\%
confidence level, and in 12 of these cases the 3-T model is not
significantly better than the 2-T model. This suggests that the poor
fit is not due to the limited number of components adopted, but
rather to other causes, perhaps residual problems in the calibration
of the instrument response. Inspection reveals in most of these
cases large residuals near the iridium edges at 5.7--5.9\AA\
associated with the coating of the Chandra mirror. A formally better
$\chi^2$ could be obtained by ignoring a narrow wavelength range
in this spectral region, with little variations of the best-fit parameters. 
For several sources, we cannot exclude also an astrophysical process such as a
departure of the plasma conditions from thermal equilibrium or
temporal and/or spatial variations of element abundances in the emitting
plasma. Recall that most of the sources are strongly variable and
characterized by important flaring events. Nonetheless, since source
variability is the norm rather than exception for ONC stars, we have
not discarded any source based on a variability criterion.

\subsection{Validation of spectral parameters}

\begin{figure}[!ht]
\plotone{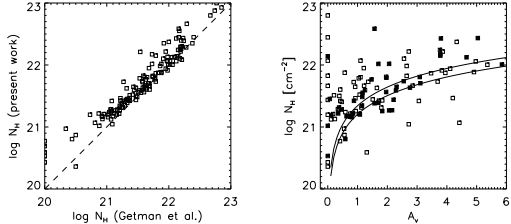}
\caption{
{\it Left:} Comparison of absorption column densities obtained by 
\citet{g+05} with those in the present work.
{\it Right:} Plot of absorption column density from COUP spectral fits
vs.\ visual absorption. The two curves give the gas-to-dust relationship
$N_{\rm H} = 1.6 \times 10^{21} A_{\rm v}$ (lower)  and
$N_{\rm H} = 2.2 \times 10^{21} A_{\rm v}$ (upper). Full symbols are
used for stars in the count-limited sample.
\label{fig:nh}}
\end{figure}

\begin{figure}[!ht]
\plotone{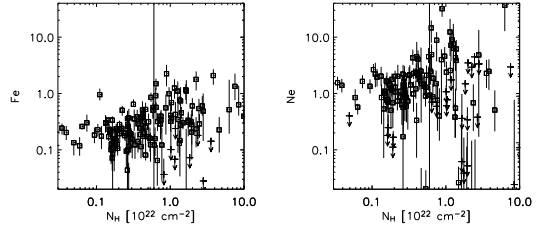}
\caption{
Scatter plots of best-fit abundance measurements
of Fe ({\it left}) and Ne ({\it right}) vs.\ H column density.
Plus and arrow symbols indicate upper limits.
The heavy vertical line is drawn at the threshold value we have
adopted to define our low-absorption sample.
\label{fig:abs}}
\end{figure}

We have estimated the uncertainty on individual quantities derived from these
highly nonlinear spectral fits using the {\it XSPEC} code,
with the criterion $\Delta \chi^2 = 2.7$ around 
$\chi^2$ minimum, corresponding to the 90\% confidence
level for one interesting model parameter at a time \citep{lmb76}.
In particular, we have evaluated uncertainties in Fe and Ne abundances,
elements with the strongest line signatures in our spectra,
by allowing temperatures, emission measures, and abundances of the four elements
with lines in the wavelength range 9--14.5\,\AA, (Ne, Mg, Fe, and
Ni; see Figure~\ref{fig:emiw}) to vary freely.  The resulting
uncertainties are mostly less than a factor of two around the fitted
value; more precisely, for half the measurements the relative error is within
30\%, and it exceeds a factor 2 only in 5\% of the cases
(see error bars plotted in Figure~\ref{fig:abs}).

We have evaluated the reliability of the derived abundances and their
uncertainties in the simulations described in Appendix A.  We find
that the overall pattern of abundances for most elements is
recovered with little bias by our analysis procedure, although some
elements (Ca, Ni, Mg and O in particular) could be vulnerable to 
systematic errors.
The {\it XSPEC} errors for Fe and Ne abundances are
sometimes smaller than indicated by the simulations (\S~\ref{sec:robust}).

One of the sources of uncertainty could be the actual emission measure
distribution vs.\ temperature in the emitting plasma. To test the
possibility that our element abundances derived from global spectral
fitting could be affected by inadequate X-ray emission models,
we have performed simulations with input emission measure distributions more
complex than simple 2-T or 3-T approximations. We then checked that
the plasma abundances are correctly recovered in spite of the
mismatch between the actual source emission measure distribution and
the adopted fitting model. For a few sample stars, we also fitted the
observed X-ray spectra with 
alternative plasma emission models, having a fixed grid of
temperatures and variable emission measures, and we have obtained abundance
measurements consistent with the results presented above, within
statistical uncertainties. These simulations and tests give us
confidence that our spectral analysis provides us with a reasonably
accurate description of the abundance patterns in the observed
coronal plasma, at least for the sources without strong interstellar
absorption (see below).

A final check on our spectral fitting is shown in
Figure~\ref{fig:nh} where the derived interstellar hydrogen column
densities ($N_{\rm H}$) are compared with values obtained by Getman
et al. (2005) for all COUP sources and with dust reddening estimated
from optical spectroscopy.  Our values are very closely correlated to
the COUP values except for a systemic offset by about 0.1 dex.  We
attribute this small discrepancy to the differences in the spectral
fitting procedure used in the COUP analysis (1-T/2-T MEKAL plasma
models rather than 2-T/3-T APEC models).  The scatter in the $N_{\rm
H} - A_V$ plot is similar to that seen in the full COUP sample.

As the $N_{\rm H}$ values in our sample exhibit a wide range from
$\sim 2 \times 10^{20}$ to $\sim 10^{23}$\,cm$^{-2}$, we
have investigated whether absorption might affect the elemental abundance
measurements. Figure~\ref{fig:abs} shows a strong increase in the
spread of Ne and Fe abundances, accompanied by larger
statistical errors, as $N_{\rm H}$ increases.
To minimize the influence of the above effects but still
keeping the sample size as large as possible, we apply a threshold
$N_{\rm H} < 6 \times 10^{21}$\,cm$^{-2}$ (corresponding to $A_{\rm
V} < 3$--4) to avoid the high scatter in Ne and other abundances due
to absorption. At this threshold, the attenuation at the Ne~{\sc x}
Ly$\alpha$ wavelength (12.13\,\AA) is about a factor 4.
There are 86 sources in our low-absorption sample, and we will focus
our attention on them in the next sections.

\section{Results}
\label{sec:res}

The best-fit spectral model parameters for the 86 stars in the
low-absorption sample are reported in Table~\ref{tab:res}. 
A subset of 35 sources in this sample have more than $10^4$ counts in
their spectra, and we will call it the count-limited subsample.
The table gives the derived absorption, plasma temperatures and 
emission measures, abundances for 9 elements, the reduced $\chi^2$ of
the fit, and the source X-ray flux in the 2--8\,keV band. 
The median values of the abundance distributions and the central 68\%
ranges are reported in Table \ref{tab:abres}, for both the
low-absorption sample and the count-limited subsample.

\subsection{Coronal temperatures and elemental abundances}
\label{sec:tha}

\begin{figure}[!ht]
\plotone{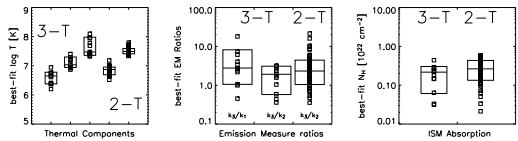}
\caption{
Box plots of temperatures, ratios of emission measures, and H column
densities, derived from 2-T and 3-T fits (see Table \ref{tab:res}). 
The upper and lower edges
of each box comprise the central 68\% of the data, the central value
is the median. Squares mark individual measurements.
\label{fig:tab}}
\end{figure}

Figure \ref{fig:tab} shows boxplots of temperatures, ratios of
emission measures, and H column densities for the low-absorption
subsample. These ONC stars are characterized by coronal plasma with
temperatures ranging from $\approx 5$\,MK to 25\,MK (median values for
the 2-T or 3-T models). The high-temperature components are dominant
in most cases with emission measures typically two times larger than
for the cool ($T < 10$\,MK) components. It is worth noting that the
thermal characteristics of our sample stars are optimal for
measuring Ne abundances, together with those of Mg and Ni, because
the emissivities of the relevant emission lines peak in the same
temperature range (see Figure~\ref{fig:emiw}).

\begin{figure}[!ht]
\plotone{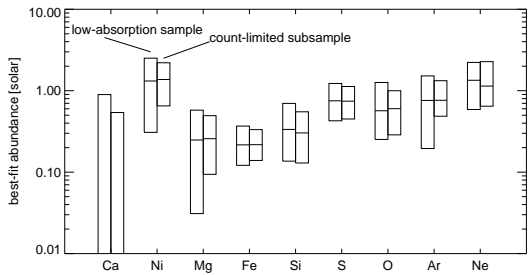}
\caption{
Box plots of the best-fit abundance distributions for each element,
sorted by increasing First Ionization Potential. The boxplot on the
left in each pair refers to the full sample of
slightly absorbed sources ($N_{\rm H} < 6 \times
10^{21}$\,cm$^{-2}$), while the boxplot on the right pertains to 
the count-limited subsample ($> 10,000$ extracted counts).
\label{fig:abdist}}
\end{figure}

Boxplots of best-fit abundance values vs.\ First Ionization
Potential (FIP) for the low-absorption sample and for the count-limited
subsample are shown in Fig.~\ref{fig:abdist}. The three values
indicated by each box (lower and upper edges, and central segment)
represent the 68\% range and the median reported in
Table \ref{tab:abres}.

The striking feature of all these plots is the systematic pattern of
abundance values vs.\ FIP: relatively low abundances with respect to
the solar photospheric composition are consistently found for
low-FIP elements Fe, Mg, and Si while elements with higher FIP are
increasingly more abundant.  Calcium and nickel abundances do not
follow this trend, and our simulations (Appendix~A) confirm the
reliability of this result, in spite of some possible systematic
error in the case of the Ca, and the sensitivity to line blending
effects of the Ni measurement.

\subsection{Reliability of the spectral analysis results}
\label{sec:robust}

Before attempting any interpretation of the results we need to
discuss their robustness against a number of possible sources of
uncertainty. We first considered the count-limited subsample comprising
the 35 sources with more than 10,000 counts,
and the subsamples of sources with 2-T or 3-T best-fit models
(74 and 12 sources, respectively). These
yield essentially the same abundance distributions as the low-absorption 
sample, although with slightly different amount of scatter. 
Hence, the results do not depend on the source strength or 
assumed plasma temperature distribution.

We performed several simulations as described in Appendix~A to
investigate a variety of other possible effects. The simulations provide
us with distributions of the best-fit abundances which take into
account: the photon counting statistics at each wavelength, the
possible cross-talk between elements with emission lines falling
within the instrument spectral resolution, and the cross-talk
between line strength and continuum level determined by the
normalization of the thermal components. These simulations show that
our procedures reliably recover true source coronal abundances.
The observed FIP pattern is
recognized self-consistently only in simulations with input models
assuming that specific pattern, and
FIP effects are not artificially introduced when they are not present.

\begin{figure}[!bh]
\plotone{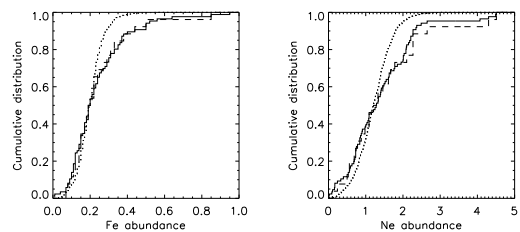}
\caption{
Cumulative distributions of Fe (left panel) and Ne
(right panel) abundances: results from simulations employing 2-T models
(dotted line) are compared with 2-T spectral fitting results for our
count-limited subsample (heavy dashed line), and with the results for the full
sample of slightly-absorbed sources (solid line).
For both elements, the observed and simulated distributions are statistically
undistinguishable at the 99\% confidence level.
\label{fig:histres}}
\end{figure}

Uncertainties of the model abundances are evaluated through the
simulations and found to have scatter similar to that seen in the
observed sample. For 2-T or 3-T source model spectra and count
rates typical of our COUP sources, our spectral analysis is able to recover
the input values within a factor of 2 
for Fe, Si, and S, and within a factor of 3 
for Ni, O, Ar, and Ne. Mg and Ca
abundances are the most uncertain by factors 5--10.

\section{Coronal abundances in the X-ray luminous ONC stars}
\label{sec:discuss}

It is important to recognize that the sample of 86 ONC stars giving
the results in Table~\ref{tab:res} and Figure~\ref{fig:abdist} is
uniquely large and homogeneous in the field of stellar X-ray
spectroscopy. Abundance measurements based on ACIS low-resolution CCD
spectra are usually impractical due to either insufficient counts or
pileup effects in high count rate stars.
Abundance measurements based on high-resolution grating X-ray
spectroscopy with Chandra and XMM-Newton are available up to now for
less than 30 late-type stars with vastly different ages and in
disparate astrophysical environments.

Given our large sample size, we can focus our attention on the
subsample of 35 sources with more than 10,000 extracted counts which
provides the most reliable abundance measurements.  We have verified
using Kolmogorov-Smirnov tests that the elemental abundance
distributions for this subsample, are statistically
indistinguishable from the distributions obtained for the sources
having between 5000 and 10000 counts. However, our entire study is
certainly biased toward the most magnetically active X-ray stars and
is limited to slightly absorbed COUP sources. The characteristics of
the more embedded Orion stellar population and of the stellar
coronae with relatively lower X-ray luminosities are not treated
here.

\begin{figure}[!ht]
\plotone{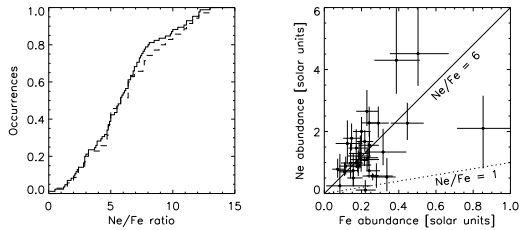}
\caption{
{\it Left:} Cumulative distributions of the Ne/Fe abundance ratios for the full
sample of slightly-absorbed sources (solid line) and for the
count-limited subsample (heavy dashed line).
{\it Right:} Scatter plot of Ne vs.\ Fe abundances, with error bars
evaluated from individual spectral fits (\S~\ref{sec:good}), for the
stars in the count-limited sample.
The two straight lines indicate different Ne/Fe abundance ratios.
\label{fig:nefe}}
\end{figure}

The next step is to establish whether our sample of 35 bright
sources is consistent with a single distribution of coronal
abundances without star-to-star variations. Figure \ref{fig:histres}
shows a comparison between the observed and simulated distributions
of abundances for Fe and Ne. The simulation here was performed
assuming a 2-T model with all parameters fixed at the median values
of the observed distributions and taking into account the actual
photon counting statistics of the observed spectra (Appendix~A).  A
Kolmogorov-Smirnov test finds that the observed distributions are
consistent with the simulations ($P \sim 10-20$\%). Similar results
are obtained for all the other elements, except for the peculiar
case of the Ca (see below).  Thus, the observed spread of abundances
for each element is compatible with being due to the uncertainties
on the measurements. There is no clear evidence for different
coronal compositions among the ONC stars in the count-limited
sample. However, we show below that marginally significant differences
between certain subsamples may be present.

Inspection of Table \ref{tab:abres} shows that the iron abundance of
the coronal plasma in our full sample of X-ray bright ONC stars is
well constrained in the range 0.12--0.37 times the solar value. The
abundances of the low-FIP elements Mg and Si are compatible with the
iron abundance, while the higher FIP elements S, O, Ar, and Ne
appear systematically higher.
Considering Fe and Ne as representative of the low-FIP and high-FIP
species, respectively, we find a median Ne/Fe abundance ratio of
6$\pm$2 (Fig.~\ref{fig:nefe} left). 
While the uncertainties on the
measurements for individual stars can be large (Fig.~\ref{fig:nefe}
right), the median value is well-measured and the possibility that
Ne/Fe=1 is confidently excluded.

Nickel and calcium abundances do not follow a simple FIP-abundance
relationship.  The relatively high abundance of the low-FIP Ni may
appear suspicious. 
The atomic database we have employed contains a large
number ($> 1600$) of L-shell Ni lines in the range 5--24 \AA,
produced by all ions from Ne-like \ion{Ni}{19} to Li-like
\ion{Ni}{26}, which form at temperatures ranging from 8\,MK to
25\,MK, hence it is quite complete in this respect.
The most prominent spectral lines are those of
\ion{Ni}{19} at 12.44, 12.66, 13.78, 14.04, and 14.08\,\AA, 
which fall close to the
important H-like and He-like Ne lines, and a cross-talk between the
two abundance parameters is possible. However, our simulations
indicate that a high Ni abundance can be correctly recovered
(Appendix~A). We conclude that the high Ni abundance looks
real, although the uncertainties may be larger than for other
elements.

 \begin{figure}[!bh]
 \plotone{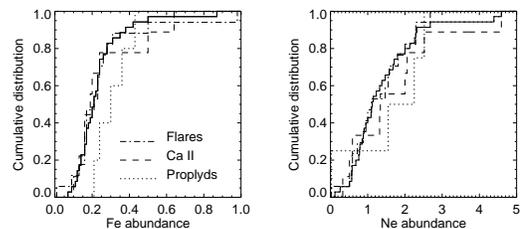}
 \caption{
 Cumulative distributions of Fe (left panel) and Ne
 (right panel) abundances for the count-limited sample (solid line) and
 for three different stellar groups: X-ray sources associated with
 proplyds (dotted line), stars with \ion{Ca}{2} in emission
 (heavy dashed line), and stars with evidence of 
 large flares (dot-dashed line).
 \label{fig:cmphist}}
 \end{figure}

The best-fit calcium abundances are zero for about 70\% of the stars
in our sample, independently from the amount of hot plasma indicated by
the best-fit model. But Ca K-shell lines are clearly visible at $\sim
3$\,\AA\ in the spectrum of
several COUP sources and give high measured Ca abundance values
(Table~\ref{tab:res}). 
Other important L-shell lines from \ion{Ca}{16}--\ion{}{18} fall in
the range 19--24\AA, i.e. at the end of the inspected ACIS band; these
lines are much weaker than the \ion{O}{7}--\ion{}{8} lines occurring in
the same spectral region, and hence they are not useful for the
determination of the Ca abundance.
Simulations performed with the Ca abundance
set to zero predict a distribution of best-fit Ca values which is
below the observed distribution, and simulations assuming a
solar Ca abundance give a higher distribution (Appendix~A). This
suggests either an intrinsic spread in Ca abundance is present in
the sample, or that some unknown systematic error in the analysis
affects calcium abundance estimates.  At present, we believe that
the reported underabundance of Ca in most sources is a solid result.

\section{Abundances as a function of other properties}
\label{sec:phys}

The above analysis indicates that the stars in our sample, chosen to
have very high X-ray luminosities ($L_{\rm x}$ in the range
$10^{29.8}$--$10^{31}$\,erg s$^{-1}$), share similar temperature distributions 
and chemical abundances. These similarities represent a major result 
of the work presented here and suggest that a single physical mechanism is
operative in the sample.

We can nonetheless investigate whether interesting subsamples behave
as the whole population of X-ray-bright ONC stars. Inspection of the
optical characteristics of our ONC sample (Table~\ref{tab:opt})
reveals that our sample includes nine stars with \ion{Ca}{2} in
emission suggesting active accretion (COUP 11, 66, 112, 141, 567,
579, 670, 801, and 1608), and five stars associated with imaged
proplyds (COUP 382, 579, 597, 758, and 1409). Seventeen stars in our
sample were studied by Favata et al. (2005) for the presence of
large flares in their COUP light curves; six (COUP 43, 141, 669,
752, 848, 1608) showed evidence of very long 
($L > 5 R_*$) flaring magnetic loops.

In Figure~\ref{fig:cmphist}, we compare the Fe and Ne abundance
distributions for our count-limited sample with abundances of these
three groups.  We find that the X-ray sources associated
with proplyds show on average higher abundances with respect 
to the sources in the count-limited sample, while the strong-\ion{Ca}{2} stars
and the stars in the flaring group 
are indistinguishable from the full sample.
However, even in the former case, the 
distributions for the stars in the subsample are not significantly 
different from those in the count-limited sample at 90\% Kolmogorov-Smirnov
confidence levels.

 \begin{figure}[!ht]
 \plotone{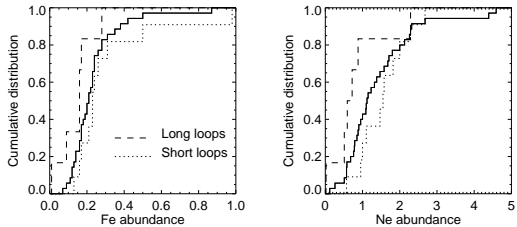}
 \caption{
 Cumulative distributions of Fe (left panel) and Ne
 (right panel) abundances for the count-limited sample (solid line) and
 for two subgroups of the X-ray sources studied by \citet{ffr+05}
 for the presence of large flares:
 stars with evidence of very long magnetic
 structures (dashed line), and the complementary sample of stars
 with shorter flaring structures (dotted line).
 \label{fig:cmphist2}}
 \end{figure}

The case of the stars caught during strong flares is particularly
interesting because time-resolved analyses of large flaring events in
active stars have indicated in many cases an apparent increase of the plasma
metallicity at the onset of the flare with respect to the 
quiescent phase \citep[and references therein]{fm03}.
We thus might expect that the COUP stars whose X-ray emission 
is dominated by strong flares could have higher Fe and Ne
abundances. We do not observe this effect in Figure~\ref{fig:cmphist}, 
where the flaring stars are considered all together.

A more intriguing case is offered by
Fig.~\ref{fig:cmphist2}, which compares the abundance distributions of
the long-loop stars with those characterized by shorter flaring
loops, as derived by \citet{ffr+05}. 
While the abundances in the short-loop flaring plasma tend to be
higher than the average, the long-loop objects show systematically 
lower abundance values (also for the other elements not shown in figure).
Kolmogorov-Smirnov tests
performed between these two distributions yield probabilities $P=9$\% (1\%) 
that the Fe (Ne) abundances are drawn from the same parent population.
These low probabilities suggests that we are indeed observing different classes
of X-ray sources in the two groups, characterized by different
chemical evolutions or different origins of the flaring plasma.
Since the statistical significance of this result is not very high,
specific time-resolved analyses of individual flaring events are
required to confirm it.

\section{Comparison with older magnetically active stars}

\begin{figure}[!ht]
 \plotone{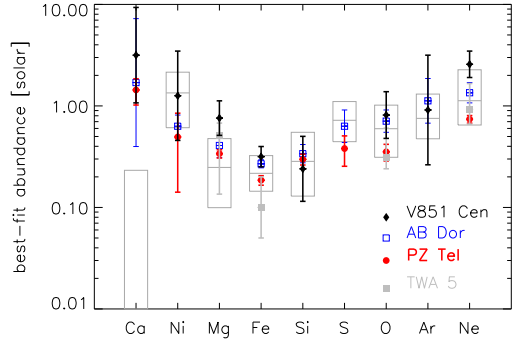}
 \caption{
 Comparison of abundances derived for the COUP sources (box plots
 of the fitting results for the count-limited
 subsample) with the abundances
 obtained from the analysis of high-resolution grating spectra of
 four active stars: TWA5, PZ\,Tel, AB\,Dor, and V851\,Cen.
 \label{fig:abcomp}}
 \end{figure}

We return to the ensemble properties of the X-ray bright,
count-limited sample of 35 ONC stars to compare with other active
stars for which high-quality X-ray spectroscopy is available.
Fig.~\ref{fig:abcomp} shows the abundances ordered by FIP derived
for our COUP sources with four comparison stars: the classical T
Tauri star TWA\,5 in the TW Hya association \citep{a+05},
the weak-line T Tauri star PZ\,Tel in the $\beta$ Pic association
\citep{a+04}, the ZAMS star AB\,Dor \citep*{smm03},
and the active binary system V851\,Cen \citep*{sfm04}.
The abundances for these four stars were derived from
high-resolution grating spectra taken with Chandra and/or
XMM-Newton, and hence with techniques more refined than the global
spectral fitting approach used here.

The similarity of the abundance patterns $vs.$\ FIP in the Orion and
older stars is striking, except for the discrepant calcium
abundances.  The X-ray bright ONC stars share with other active
stars a characteristic Ne/Fe abundance ratio several times the
solar ratio \citep{dbk+01,g04}. This behaviour is often attributed to an underabundance of
low-FIP elements with respect to high-FIP elements, i.e. to the so called
"inverse FIP effect" \citep{b+01}.
Such a behaviour was recently observed
also in young, weak-line T~Tauri stars \citep{a+04,a+05},
and now we find it in the X-ray bright ONC stars.

In recent years, the number of coronal sources with available
abundance determinations has increased steadly, but we are far from
a clear assessment of the phenomelogy. In fact, a variety of
abundance patterns has been observed, with more or less pronounced
deviations from both the classical solar FIP effect and the inverse
FIP effect in its original version.  For example, the four
comparison stars in Fig.~\ref{fig:abcomp} show the inverse-FIP
effect between iron and neon but they are all characterized by
relatively high abundances for the low-FIP elements Ca, Ni, and Mg.
Some star-to-star differences in the abundance patterns appear to be
linked to the stellar activity level \citep{ags+03,g04,gdb+06},
but again with striking exceptions. 
\citet{wl06} have recently reported the case of
the binary 70\,Oph where the primary shows a prominent solar-like
FIP effect while the secondary has no FIP bias or possibly a weak
inverse FIP effect, in spite of the similarity between the two stars
in all other respects.

\section{Revised treatment of standard abundances}

Part of the confusion is likely due to our ignorance of true stellar
photospheric abundances of the magnetically active stars, which are
usually {\em assumed} to be solar, or at least with the same ratios
as in the solar photosphere. When proper stellar abundance
measurements are employed, the abundance vs.\ FIP pattern is no
longer very clear \citep{sfm04}.

\subsection{Orion photospheric and nebular abundances}
\label{sec:Orion_abun}

In the case of the ONC, an assessment of the photospheric
composition is available only for a handful of stars. The chemical
evolution of the Orion association was studied by \citet{cl92,cl94}
who derived photospheric CNO, Si and Fe abundances for
18 main-sequence B stars. One of the results of these early works
was that the spread in O and Si abundances was larger than expected
based on the measurement uncertainties.  This is thought to indicate
a real spread among stars of different ages, caused by
self-enrichment of the nebula as a consequence of supernova
explosions within the Orion association. However, the result could
also be affected by systematic errors in the analyses, due to
approximations in the adopted NLTE model atmospheres and
line-blanketing effects. In fact, detailed calculations of the
oxygen abundances for 3 B stars in Orion recently presented by
\citet{sd+06} yield values lower by $\approx 0.2$ dex
with respect to those of \citet{cl94} for the same stars.

\begin{figure*}[!th]
\plotone{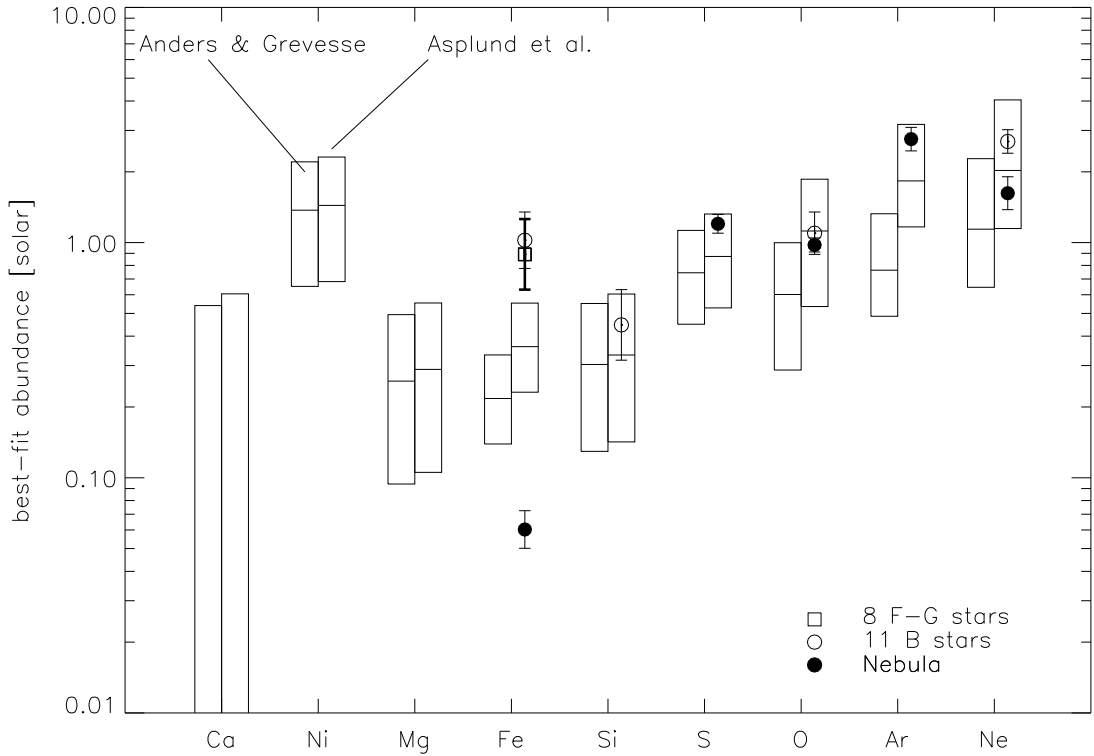}
\caption{
Comparison between the abundances derived for the COUP sources in the
count-limited subsample -- scaled by two different sets of reference
solar values as indicated -- with the abundances
obtained for 11 B-type stars in the Trapezium \citep{cl94,chl06}, 8 F-G stars
\citep{csl98}, and for the Orion nebula \citep{e+05}.
\label{fig:compneb}}
\end{figure*}

\citet{chl06} report NLTE Ne
abundances for 11 B-type stellar members of Orion and found a
homogeneous abundance of neon, $A({\rm Ne}) = 8.27 \pm 0.05$, and
oxygen, $A({\rm O}) = 8.70 \pm 0.09$ (in a log scale where $A({\rm H}) =
12$). For the same sample,
we have computed average Si and Fe abundances from the measurements of
\citet{cl94}: $A({\rm Si}) = 7.16 \pm 0.15$, and $A({\rm Fe}) = 7.46 \pm 0.12$. 
These values will be used in the next section.

For later-type stars, \citet{csl98} determined NLTE oxygen and
LTE Fe abundances from optical spectroscopy of 9 pre-main-sequence F
and G Orion stars.  \citet{cs05} report a study of
fluorine, C and O abundances in 3 Orion K-M dwarfs. These works
indicate that the solar-type stars of the Orion association all have
the same Fe abundance:  $A({\rm Fe}) = 8.40 \pm 0.15$.
In contrast, the oxygen abundance appears
to vary from star to star with a large spread ($A({\rm O}) = 8.94
\pm 0.36$ for the full sample), and we will not consider these
measurements in the following.

Finally, we consider the composition of the Orion Nebula, which is
the brightest and nearest Galactic \ion{H}{2} region in the sky.
\citet{e+05} present echelle spectrophotometry of a region
S-W of $\theta^1$\,Ori C and derive abundances of several ionic
species, including \ion{Ne}{1} and \ion{Ne}{2}, from
collisionally-excited lines or recombination lines.  Their analysis
takes into account spatial variations of the temperature structure
of the nebula and, applying ionization and dust-depletion
corrections, obtain abundances of several gas-phase elements.

\subsection{Revised solar abundances}

\citet*{ags05} present a detailed 3-D
hydrodynamic modeling of the solar atmosphere and find that the
abundances of many elements (including C, N, O, Ar, Ne, and Fe) need
to be revised downward by factors of 1.5--2.4 from the widely-used
compilation of \citet{ag89}. But the new solar
composition implies lower opacities and produces a severe
inconsistency between the standard solar interior model and precise
helioseismology measurements \citep{ab05,bbs05}.
One solution to this conundrum is to revise upward
the poorly-known Ne abundance in the Sun, bringing it closer to the
values often measured in stellar coronae.  From high-resolution
X-ray spectroscopy of several active late-type stars, \citet{dt05}
showed that the coronal Ne/O abundance ratio is, on average,
a factor 2.7 times higher than the solar value recommended by
\citet{ags05}. A similar result was obtained by \citet{chl06},
who presented measurements of the photospheric Ne abundance
in a sample of B-type stars in Orion, and obtained a Ne/O abundance
ratio a factor 2.5 higher than the most recent solar value.

We also find the COUP sources in the count-limited sample we have
obtained a median Ne/O ratio of 0.33, i.e. a factor 2.2 higher than
the \citet*{ags05} value. 
However, this ratio
suffers large scatter in individual objects (the central 68\% of the
data span the range 0.11--1.18), possibly because 
oxygen measurements are the most affected by uncertainties in the amount
of absorption and in the amount of low-temperature plasma (see
Appendix~A).

\subsection{ONC coronal abundances with revised standard abundances}

The boxes in Fig.~\ref{fig:compneb} shows our Orion coronal
abundances inferred from the COUP X-ray spectra with respect to the
traditional \citet{ag89} and revised \citet{ags05}
solar abundances. We have not adjusted the solar neon
abundance as suggested by the work of \citet{dt05}. The boxplot shows
that the inverse-FIP abundance pattern for our ONC stars is still
present, and is even slightly more pronounced, with the \citet{ags05}
solar abundances.

The points with error bars in Figure~\ref{fig:compneb} show average
stellar photospheric abundances and nebular abundances (not corrected
for dust-locking effects) as described in \S~\ref{sec:Orion_abun},
scaled by the \citet{ags05} revised solar
abundances (a similar scaling is made by \citep{e+05}). Our
COUP coronal abundances for the high-FIP elements S, O, Ar, and Ne
are very similar to those of the nebula, and also show good
agreement with the stellar photospheric values for Si, O, and Ne.
Thus, while we do find a strong inverse FIP effect with respect to
solar elemental abundances, the effect disappears when Orion
photospheric and nebular abundances are considered.

However, discrepancies are found in the iron abundances which
appear significantly lower in the X-ray coronal plasma than in the
stellar photospheres. The very low Fe abundance found for the
gaseous nebula can be attributed to heavy depletion into grains. 
If the value derived for the B-type and F-G stars is
indeed representative of the iron abundance in the photospheres of
all the late-type ONC stars, Figure~\ref{fig:compneb} suggests that
the coronae of these stars are depleted in iron by a factor 1.5--3.

\section{Summary and conclusions}
\label{sec:concl}

\begin{itemize}
\item
The coronal temperatures and elemental abundance pattern in X-ray
luminous ONC stars is remarkably similar to that found from the
analysis of high-resolution grating spectra of older magnetically
active stars.  Hence accretion or the presence of circumstellar
disks does not appear to affect the X-ray production mechanism or
plasma. The abundance of calcium is a possible exception: it appears
to be extremely low in about 70\% of the ONC stars we have studied.
However, it is difficult to reliably measure low calcium abundances 
with the available CCD spectra.

\item
Comparison of the observed abundance distributions among different
stars with simulated distributions indicate that all stars may
actually have the same abundance values, i.e. the abundance spread
for each element is compatible with the statistical uncertainties.
Nonetheless, our results also suggest possible systematic differences
between the abundance distributions for selected subsamples
(e.g. those of the sources with short $vs.$ long flaring
magnetic structures), which require a specific time-resolved spectral 
analysis to be confirmed.

\item
The ensemble properties of the COUP X-ray brightest sources confirm
the low metallicity of the coronal plasma with respect to the solar
photospheric value: the median Fe abundance is $\approx 0.2$ times
the \citet{ag89} value, or $\approx 0.3$ times the most
recent determination by \citet{ags05}. At the same time, the
Ne/Fe abundance ratio is significantly higher than the solar one,
with a median value $\approx$ 5--7, depending on the assumed set of
solar abundances.

\item
The X-ray brightest COUP sources show a clear pattern of abundances
vs.\ FIP. Extensive simulations make us confident about the
robustness of this result. If the solar photospheric abundances are
adopted for reference, the low FIP elements (Mg, Fe, and Si) appear
to have similar low abundances, 0.2--0.3 times the solar values,
while Ni and the high-FIP elements (S, O, Ar, and Ne) appear to have
higher abundances, 1--2 times the solar ones.   However, comparison
with abundance measurements obtained by means of optical
spectroscopy of members of the Orion association indicates a good
agreement between photospheric and coronal abundances for Si, O, and
Ne, while iron is significantly depleted in the X-ray emitting
plasma with respect to the stellar photospheres, by about a factor
of 3. We conclude that there is no clear FIP-related behaviour of the
hot plasma abundances in the X-ray bright Orion stars, when proper
stellar photospheric abundances are taken into account.

\end{itemize}

\acknowledgements{AM, EF, GM, SS acknowledge partial support from
Ministero dell'Universit\'a e della Ricerca Scientifica, and from
ASI/INAF contract I/023/05/0. EDF and KVG are supported by
$Chandra$ grant GO3-4009A and the $Chandra$ ACIS Team contract
NAS8-38252.}

\newpage
\appendix
\section{Simulations to validate spectral modeling}

We have performed several simulations, each with 1000 realizations,
based on known 2-T, 3-T, or multi-temperature input models and
different elemental abundance distributions. They are employed to
assess realistic scatter of best-fit parameters, in particular on
the abundances, and to test for bias in the fitted parameters.
We have also performed specific simulations in which the photon counting
statistics of the observed spectra is taken into account.
The selected simulations are tailored for
comparison with results based on our count-limited sample with more
than $10^4$ total extracted counts. All simulations use the same
ancillary response file (i.e. instrument effective area) and
response matrix belonging to a real COUP source near the center of
the ACIS field of view. The simulated spectra were rebinned as the
actual data (\S~\ref{sec:method}) and the spectral fitting was
performed on the same fixed energy range (0.5--8\,keV) with the same
{\it XSPEC} procedures. The background spectrum associated with the
same source was used as a template in all the cases; it contributes
$0.3$\% of the total source+background counts and thus has
negligible effect on the results. In all simulations, we assume
hydrogen column density in the ISM photoelectric absorption model
component of $N_{\rm H} = 3 \times 10^{21}$\,cm$^{-2}$, which is near the
median value found for the COUP sources in our low-absorption sample
(\S~\ref{sec:tha}).

The simulations described below are sorted by increasing complexity,
so to explore different sources of uncertainty. For each simulation,
we state the issue we have tested and we show the relevant results.
These simulations validate both our ability to recover the correct
abundance pattern from the analysis of ACIS spectra, and
that the observed pattern does not arise in a spurious fashion 
by our analysis process.

The first simulation assumes a 3-T input model having temperatures,
emission measures, and abundances set to values near the median of
the distributions obtained for our sources in the low-absorption
subsample. The three plasma components have $kT_1
= 0.4$\,keV, $kT_2 = 0.8$\,keV, and $kT_3 = 2.6$\,keV ($\log T =$
6.7, 7.0, and 7.5\,K), and emission measure ratios $EM_2/EM_1 = 1.5$
and $EM_3/EM_1 = 3.5$. All simulated spectra here have 36,000 counts
before applying Poisson noise, which is near the median value of the
spectra which required 3-T best-fit models. Elemental abundances
were fixed to the median values determined by fitting the sources in
our count-limited sample; the Ca abundance, in particular, was set
to zero.

\begin{figure}
\plotone{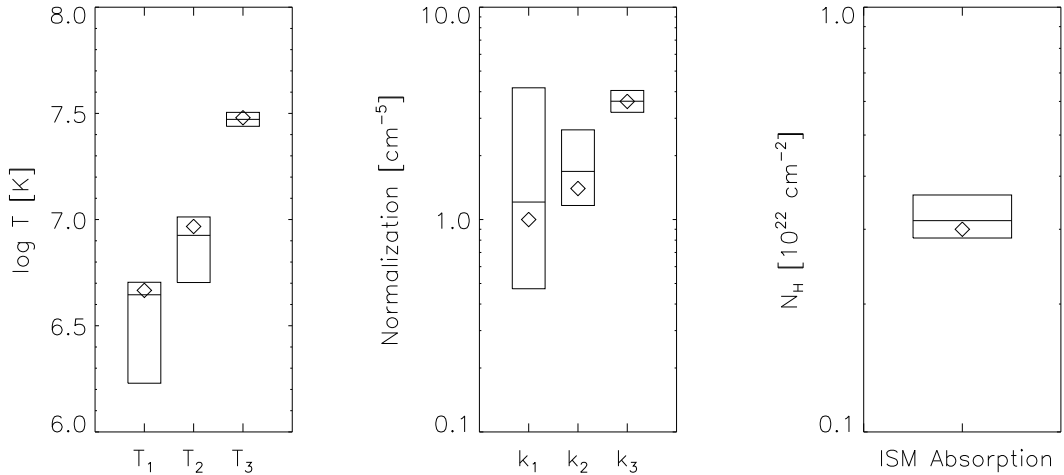}

\caption{Box plots of the best-fit temperatures (left) and
normalizations (right) derived by fitting 1000 simulated 3-T model
spectra, each with 36,000 (within Poisson statistics). The upper and
lower edges of each box comprise the central 68\% of the data, the
central value is the median. Diamonds mark the values in the input
model. \label{fig:sim6}}
\end{figure}

\begin{figure}[!ht]
\plotone{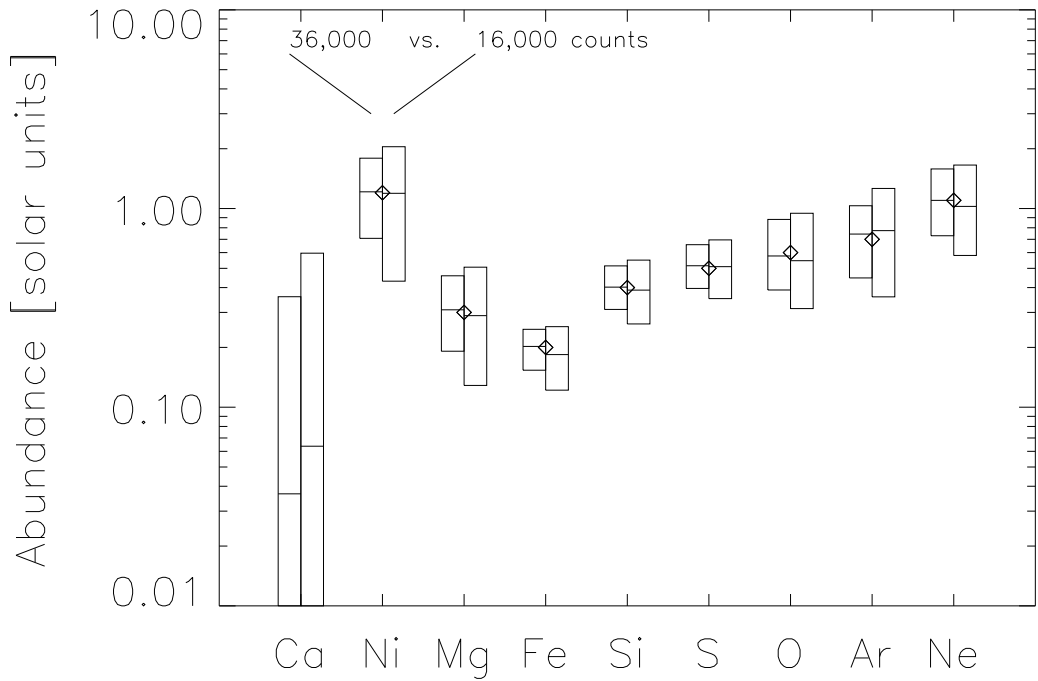}

\caption{ Box plots of the best-fit abundances derived by fitting
1000 simulated 3-T model spectra for sources with 36,000 and 16,000
counts to illustrate the effects of signal strength. Diamonds mark
the values in the input model, and the meaning of the boxes is the
same as in Fig.~\ref{fig:sim6}. \label{fig:sim67}}
\end{figure}

Figure~\ref{fig:sim6} shows the distributions of temperatures and
volume emission measures derived by fitting the simulated spectra,
while Fig.~\ref{fig:sim67} shows the distributions of the
abundances, sorted by First Ionization Potential (FIP) of the
relevant elements. The boxplots indicate the range covered by the
central 68\% of the values. 
Since the simulation is based on a perfect alignment of the
input and fitted spectral model, the scatter in parameter values serves as a
reference for "the best we can do" with Chandra/ACIS spectra.

Figure~\ref{fig:sim67} also shows the distributions of abundances
derived from simulations of 3-T spectra with 16,000
counts, which is the median for all the sources in our count-limited
sample. As expected, the width of the distributions is slightly
larger for all elements than seen in simulations with 36,000 counts.
Simulations based on 2-T
rather than 3-T input spectral models and simulations using sources
with the same distribution of counts as in our sample give results very
similar to those in Figure~\ref{fig:sim67};
little additional spread in the derived abundance
distributions is introduced by the different source model spectrum or 
by the photon counting statistics of the real COUP dataset.  

\begin{figure}[!ht]
\plottwo{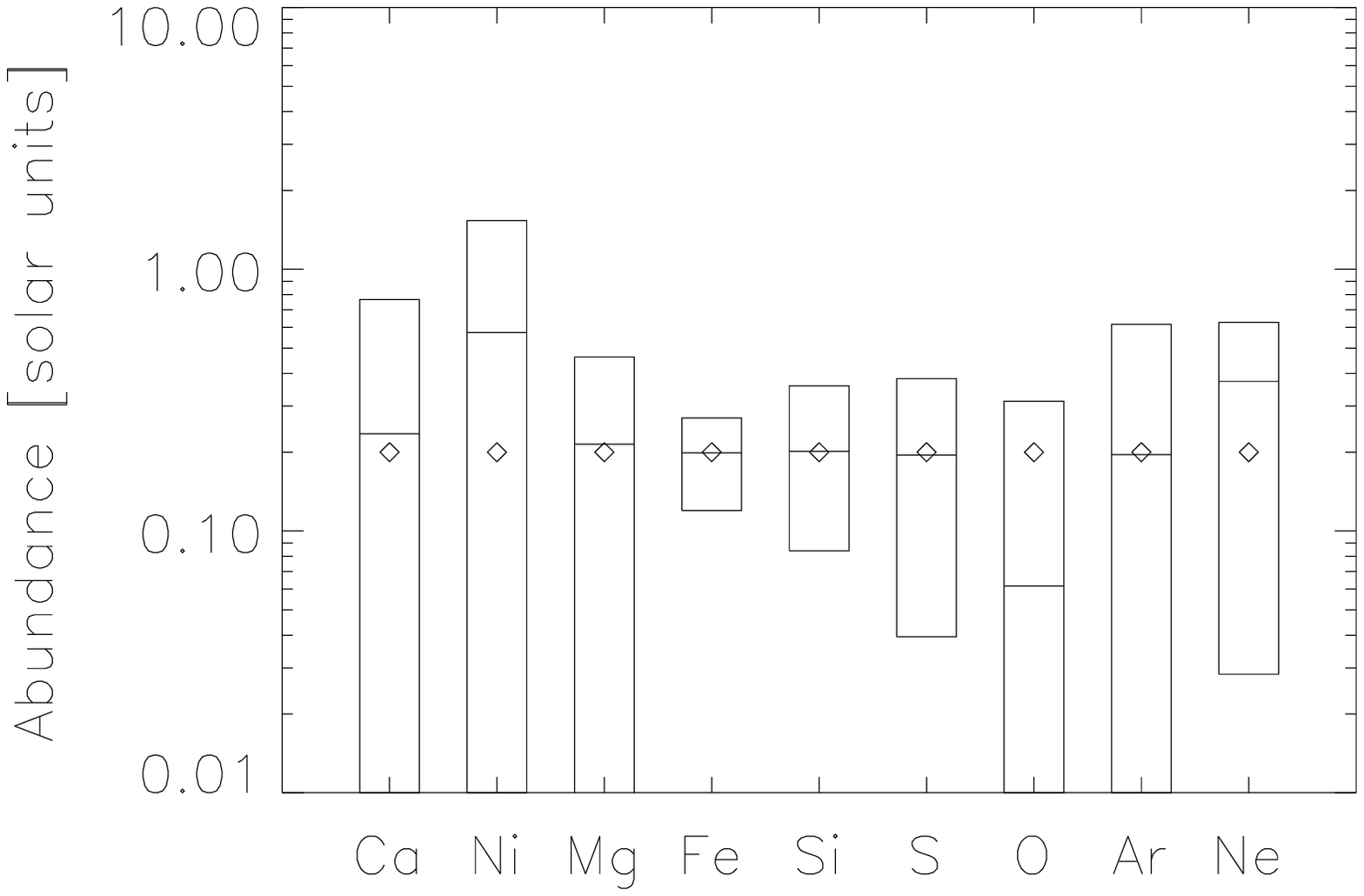}{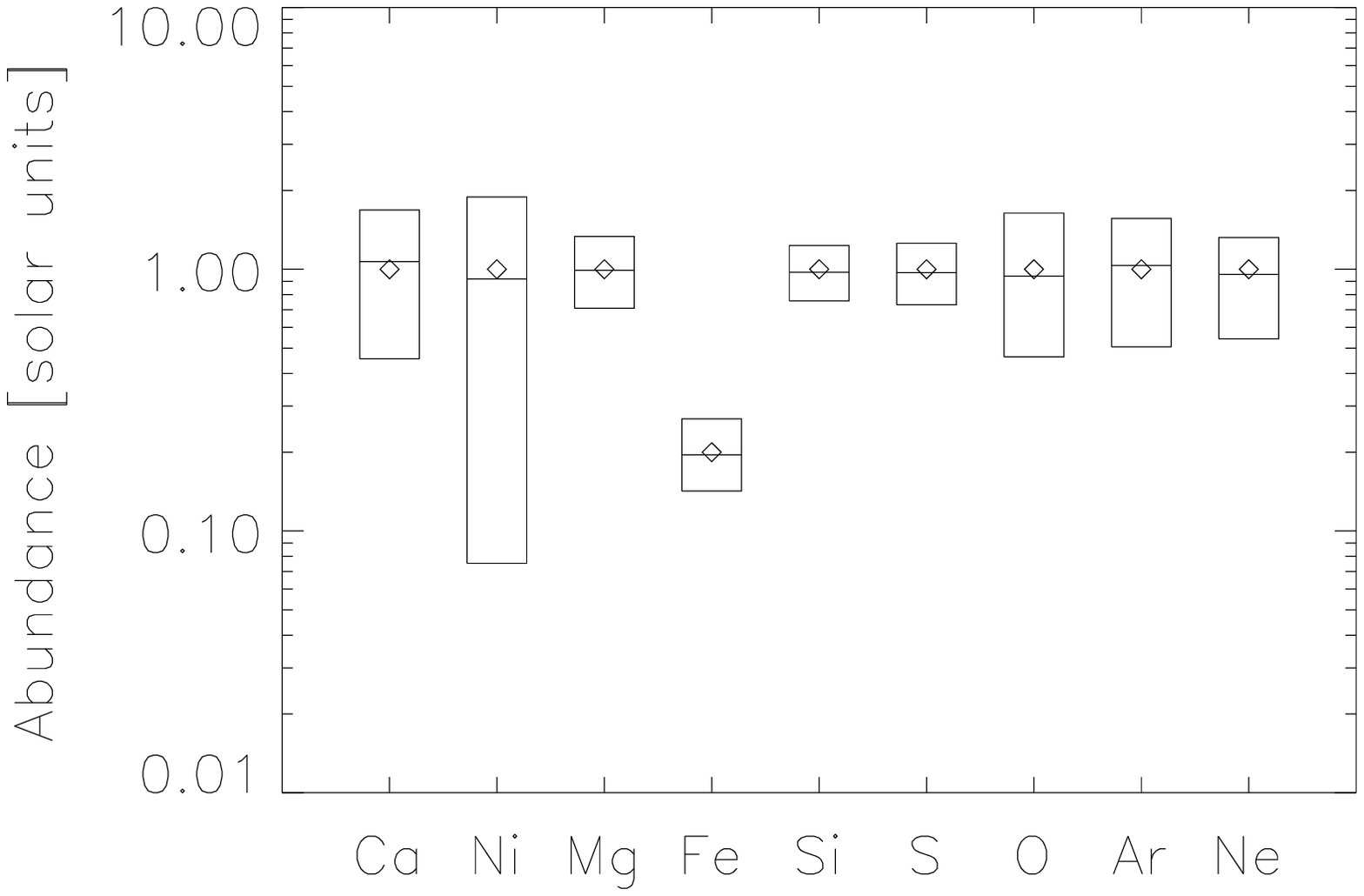}

\caption{Box plots of the best-fit abundances derived by fitting
1000 simulated 2-T model spectra with different input abundance
patterns: (left) 0.2$\times$ solar abundances; and (right)
1.0$\times$ solar abundances with low Fe abundance. Symbols and
other details as in Fig.~\ref{fig:sim6}. \label{fig:sim10}}
\end{figure}

In all cases, we see very little bias in the derived spectral
parameters; that is, the median values of the distributions lie close to
(usually within $\pm$10\% of) the input values.
Temperature estimates become increasingly inaccurate for the lower
temperature components ($\log T < 7.0$~K, which also have associated
emission measures lower than for the high-temperature component). 
Considering the results of all simulations, we find that
Fe, Si and S abundances are generally
accurate within 40--80\% relative errors, while Ni, O, Ar and Ne
have somewhat lower accuracy by factors 1.8--2.8; Mg shows the
largest scatter, with uncertainties up to a factor 10 in simulations
with 2-T models and low ($\approx 14,500$ counts) photon counting
statistics;
in the case of the Ca, whose input abundance was assumed to be zero, the
statistical fluctuations make the best-fit result the most uncertain
with any value between 0 and 0.8 acceptable.

For the cases of Fe and Ne, we have verified that the uncertainties
indicated by the simulations are slightly larger
(by factors 1.2--1.4, on average) than the XSPEC errors, 
computed at the 90\% confidence level for single parameter.
However, the results presented here show that the
uncertainties on the best-fit abundances for all elements are
sufficiently small to recover the input abundance pattern vs.\ FIP.
For example, we ascertain that the Ne abundances exceed those
of Fe with very high degree of confidence.  The scatter
on abundance ratios with iron is even lower than on individual
abundances, because all abundance measurements are correlated with the
iron one to a certain degree\footnote{This is due to the common 
inverse proportionality between abundances and
plasma emission measure while the source count rate is fixed.}:
in fact, the apparent abundance ratio Ne/Fe$\sim$6 is affected by
$\approx 30$\% uncertainty, according to our simulations.

Figure~\ref{fig:sim10} explores the possibility that the
interactions between parameters might create the observed FIP
abundance effect as a form of systemic bias in the fitting
procedure. Here we show the abundances emerging from 2-T input
models assuming 0.2 solar abundances for all elements.  No trend is
evident in the reconstructed abundance pattern. A similar result is
found for input models with 1.0 solar abundances, exhibiting less
scatter due to the stronger emission lines. Note, in particular,
that an high abundance of Ca could be correctly determined if it
were present, but little could be inferred for the Ni
abundance in this situation (quite different from the case of the COUP
sources), due to the strong blends with lines from the other elements.

\begin{figure}
\vbox{ \plotone{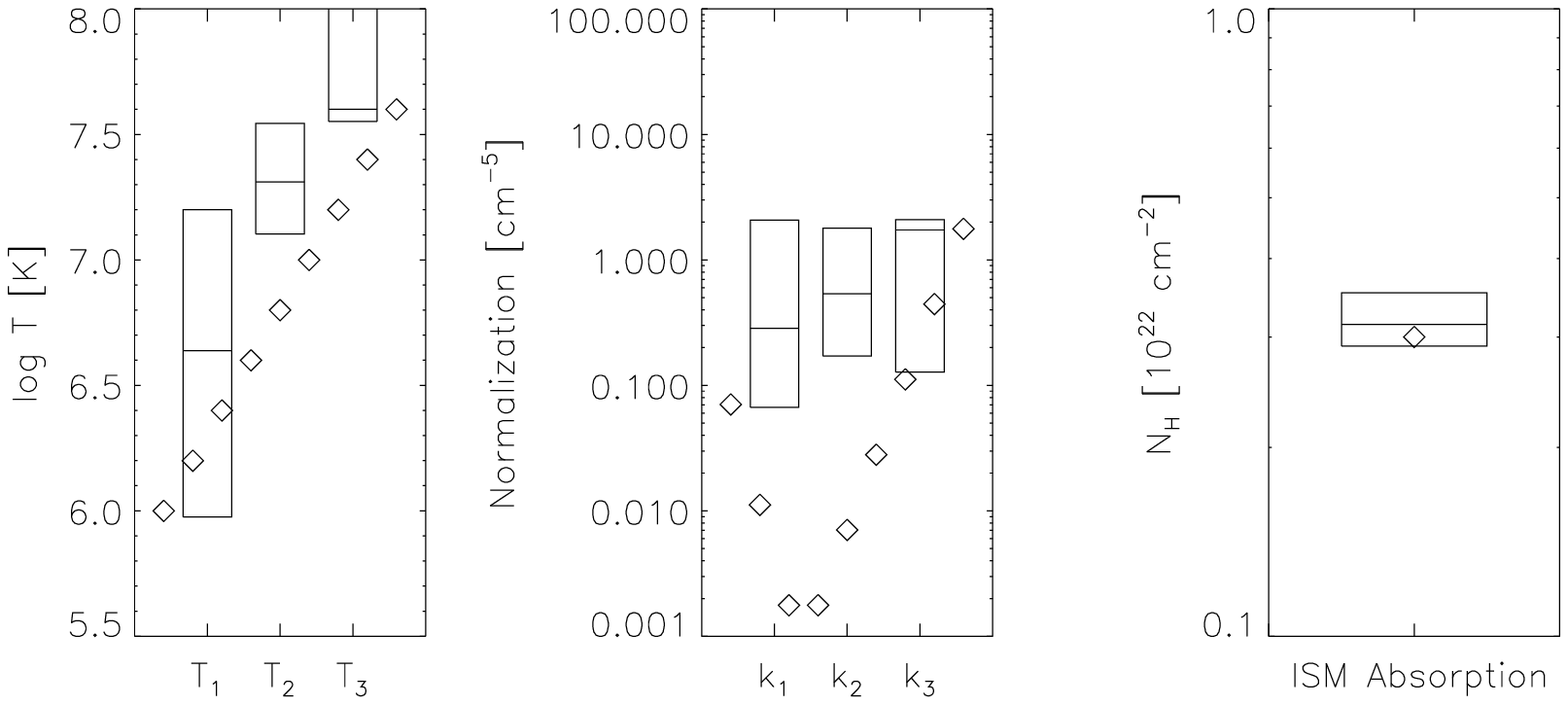} \plotone{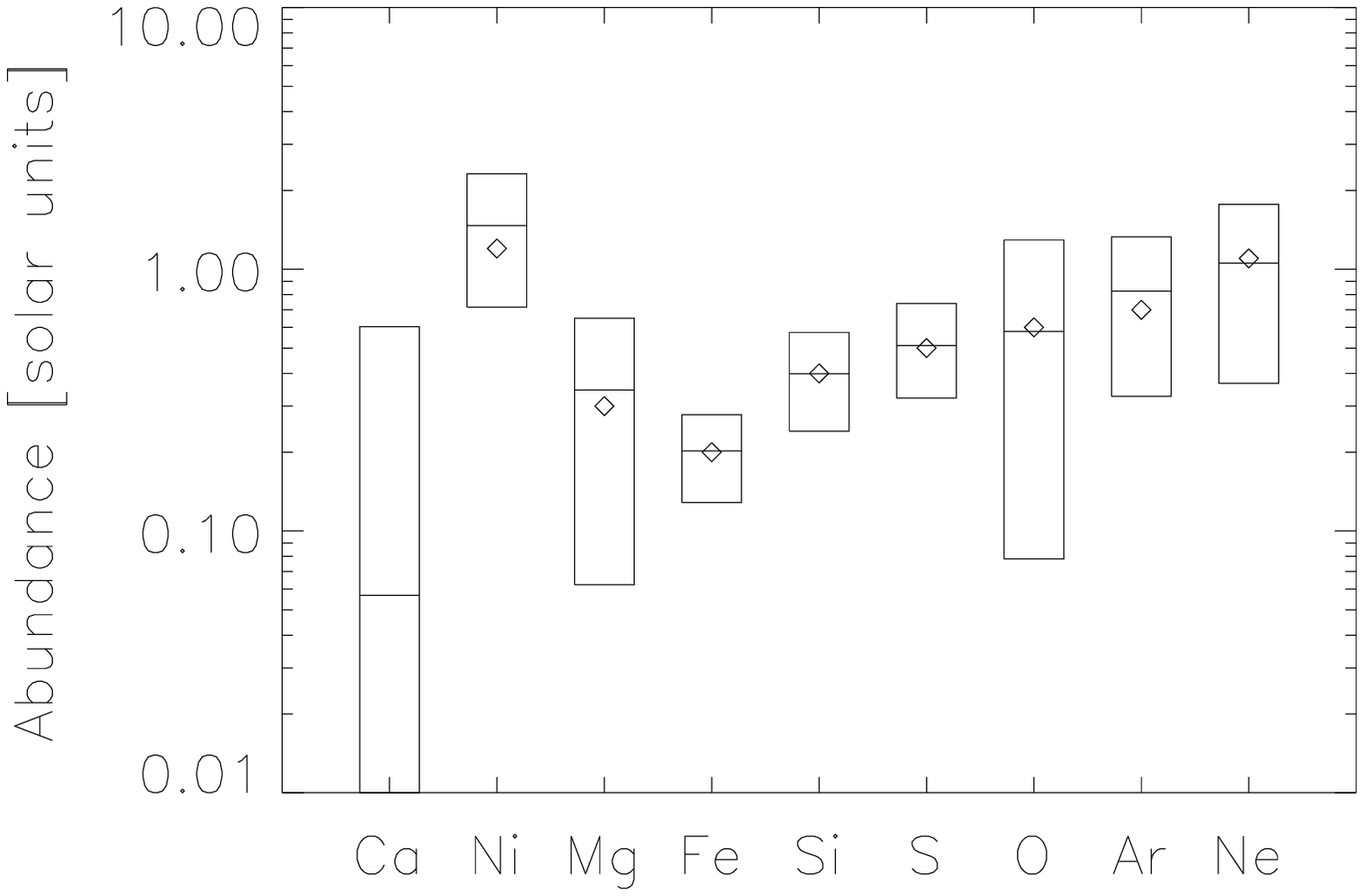}}

\caption{ Boxplots of temperatures, normalizations, H column
densities, and abundances obtained by simulating 1000 spectra with
an underlying V-shaped emission measure distribution vs.\
temperature, and fitted with 3-T models. In each panel, diamonds
show the pattern of input parameter values. \label{fig:sim12}}
\end{figure}

Our final simulations test our ability to model more complex thermal
distributions.  Simple 2-T or 3-T models are only approximations to
the actual thermal structuring of real coronae that must have continuous
distributions of emission measure $vs.$ temperature.
\citet{sn04} warn that coronal abundances may be inaccurately
estimated without continuous temperature distribution models, but this
caveat applies to the analysis of high-resolution grating spectra.
Fig.~\ref{fig:sim12} shows a representative simulation in which we
assumed a V-shaped emission measure distribution over the
temperature range $\log T = 6.0$--7.6\,K, with a minimum around
$\log T = 6.5$\,K, resembling distributions determined from
high-resolution X-ray spectroscopy of very active stars (e.g.,
Sanz-Forcada, Brickhouse \& Dupree 2003). The results of this
simulation suggest that, although the 3-T model can not effectively
recover the actual emission measure distribution, the input
pattern of abundances can still be reliably measured. Note, in
particular, that the inferred Ne is sometimes underestimated but is
not significantly overestimated with respect to the input value.
Distributions more skewed toward low values are those of the O and Mg
abundances, but the median is always quite close to
the input value.

\newpage

\LongTables
\input{tab1}
\input{tab2}
\input{tab3}

\end{document}

%% file: tab1.tex
\begin{deluxetable}{rccccrrccrrrr}
\tablecaption{Properties of sample stars 
\label{tab:opt}}
\tabletypesize{\footnotesize}
\tablewidth{0pt}
\tablecolumns{13}
\tablehead{
\colhead{COUP} & \colhead{} & \colhead{$M$} & \colhead{$\log t$} &
\colhead{$A_{\rm v}$} & \colhead{$\Delta (I$-$K)$} & \colhead{EW(Ca)} & 
\colhead{$V$} & \colhead{$I$} & \colhead{$J$} &
\colhead{$H$} & \colhead{$K_{\rm s}$} & \colhead{$L$} \\ 
\colhead{ID} & \colhead{Sp Type} & \colhead{($M_\odot$)} & \colhead{(yr)} &
\colhead{(mag)} & \colhead{(mag)} & \colhead{(\AA)} & 
\colhead{(mag)} & \colhead{(mag)} & \colhead{(mag)} &
\colhead{(mag)} & \colhead{(mag)} & \colhead{(mag)}
}
\startdata
\multicolumn{13}{c}{Low-absorption sample} \\
\hline
7 & K1-K4 & 2.12 & 5.55 & 0.75 & -0.02 & \nodata & 11.38 & 9.89 & 8.85 & 8.10 & 7.95 & \nodata \\
9 & K0-K3 & 2.11 & 6.49 & 0.88 & 0.10 & \nodata & 12.39 & 11.12 & 10.22 & 9.65 & 9.46 & \nodata \\
11 & K1e-K7 & 0.69 & 5.74 & 0.42 & 1.37 & -14.6 & 13.40 & 11.65 & 10.53 & 9.46 & 8.60 & \nodata \\
23 & K2 & 2.17 & 6.18 & 1.57 & 0.09 & \nodata & 12.72 & 11.11 & 10.01 & 9.33 & 9.09 & \nodata \\
27 & M0 & 0.53 & 6.23 & 0.94 & 0.42 & 1.8 & 15.77 & 13.61 & 12.16 & 11.37 & 11.05 & \nodata \\
28 & M0 & 0.53 & 6.01 & 0.63 & 0.30 & 1.6 & 14.95 & 12.91 & 11.53 & 10.84 & 10.53 & \nodata \\
43 & M1 (SB2) & 0.40 & 5.85 & 1.36 & 0.50 & 1.4 & 15.57 & 13.06 & 11.23 & 10.38 & 10.08 & \nodata \\
62 & K2 & 1.52 & 6.84 & 3.06 & 1.28 & 0.0 & 15.75 & 13.56 & 11.23 & 10.21 & 9.53 & \nodata \\
66 & M3.5e & 0.24 & 6.05 & 0.59 & 1.05 & -2.8 & 17.22 & 14.28 & 12.13 & 11.20 & 10.63 & \nodata \\
67 & M2.5 & 0.29 & 4.52 & 1.13 & 0.32 & 0.0 & 15.46 & 12.64 & 10.85 & 9.97 & 9.62 & \nodata \\
71 & M1.5 & 0.37 & 6.10 & 0.03 & 0.26 & 1.6 & 15.29 & 13.21 & 11.88 & 11.19 & 11.01 & \nodata \\
101 & M4.5 & 0.16 & 5.68 & 0.35 & 0.15 & 2.9 & 18.19 & 14.93 & 13.09 & 12.45 & 12.05 & \nodata \\
108 & M1.5 & 0.37 & 6.09 & 0.21 & 0.71 & 1.5 & 15.41 & 13.26 & 11.66 & 10.82 & 10.54 & \nodata \\
112 & M2e & 0.33 & 6.25 & 0.39 & 0.63 & -0.7 & 16.39 & 14.01 & 12.49 & 11.64 & 11.29 & \nodata \\
113 & A7(?) & 2.20 & 6.73 & 4.20 & 0.11 & \nodata & 13.55 & 11.73 & 10.32 & 9.65 & 9.37 & \nodata \\
139 & M2 & 0.33 & 6.20 & 0.18 & 0.71 & 0.9 & 16.03 & 13.73 & 12.12 & 11.28 & 10.91 & \nodata \\
141 & B9-A1 & 2.11 & 6.47 & 1.83 & 0.43 & -17.8 & 12.94 & 11.36 & 10.23 & 9.42 & 8.99 & \nodata \\
150 & M2.5 & 0.29 & 5.00 & 1.51 & 0.44 & 1.5 & 16.38 & 13.41 & 11.48 & 10.61 & 10.29 & \nodata \\
152 & \nodata   & \nodata & \nodata & \nodata & \nodata & \nodata & 14.91 & 12.83 & 11.38 & 10.60 & 10.30 & \nodata \\
173 & M1.5 & 0.37 & 5.60 & 0.47 & 0.67 & 1.0 & 14.92 & 12.67 & 10.91 & 10.15 & 9.86 & \nodata \\
177 & K5 & 1.19 & 6.36 & 2.85 & 0.48 & 2.0 & 16.06 & 13.64 & 11.54 & 10.52 & 10.10 & \nodata \\
188 & K1-K2 & 2.16 & 6.25 & 2.21 & 0.23 & \nodata & 13.56 & 11.70 & 10.33 & 9.51 & 9.22 & \nodata \\
202 & M1.5-M4 & 0.37 & 5.62 & 0.47 & 0.38 & 1.5 & 14.98 & 12.73 & 11.24 & 10.44 & 10.12 & \nodata \\
205 & M2 & 0.33 & 6.13 & 0.34 & 0.64 & 1.2 & 15.94 & 13.58 & 12.08 & 11.19 & 10.90 & \nodata \\
270 & M1 & 0.41 & 5.98 & 0.24 & 0.32 & 1.5 & 14.93 & 12.86 & 11.45 & 10.68 & 10.38 & 10.34 \\
328 & K1-K6 & 1.72 & 6.48 & 0.85 & 0.08 & 1.8 & 13.20 & 11.78 & 10.76 & 10.05 & 9.87 & \nodata \\
343 & K4-M0 & 0.59 & 5.76 & 0.12 & 0.52 & 1.9 & 13.47 & 11.71 & 10.46 & 9.66 & 9.39 & 9.25 \\
382 & K2-M2 & 0.69 & 6.06 & 0.42 & 0.81 & 0.0 & 14.28 & 12.53 & 11.24 & 10.39 & 9.94 & 8.90 \\
387 & K0-M0 & 2.34 & 6.43 & 1.70 & 1.00 & \nodata & 12.69 & 11.16 & 9.70 & 8.83 & 8.26 & \nodata \\
417 & M1 & 0.41 & 6.24 & 0.37 & \nodata & \nodata & 15.82 & 13.70 & 12.12 & 11.30 & 11.04 & \nodata \\
431 & G0-K0 & 2.61 & 6.45 & 3.11 & -0.01 & \nodata & 12.79 & 10.94 & 9.54 & 8.86 & 8.63 & \nodata \\
459 & M0.5 & 0.27 & 5.00 & 0.00 & \nodata & \nodata & 14.32 & 12.45 & 11.07 & 10.34 & 10.08 & \nodata \\
470 & K1-M0 & 0.52 & 5.95 & 0.81 & 1.08 & 1.5 & 15.00 & 12.89 & 10.72 & 9.91 & 9.60 & 8.87 \\
567 & F8-K5e & 1.20 & 6.02 & 0.38 & 0.80 & -3.5 & 12.94 & 11.48 & 10.18 & 9.26 & 8.62 & \nodata \\
579 & K2e-M4 & 0.33 & 5.16 & 0.00 & 1.31 & -17.4 & 14.40 & 12.30 & 10.80 & 9.59 & 8.78 & 8.13 \\
597 & late-G & 1.49 & 7.06 & 2.69 & \nodata & 4.5 & 14.44 & 12.69 & 11.47 & 10.61 & 10.06 & 9.34 \\
600 & M3.1 & 0.26 & 4.34 & 1.83 & 0.93 & \nodata & 16.30 & 13.04 & 11.13 & 10.18 & 9.26 & 9.66 \\
648 & K3-M1.5 & 0.72 & 5.30 & 2.29 & \nodata & \nodata & 14.58 & 12.10 & 10.44 & 9.53 & 9.14 & 9.16 \\
669 & K3-K4 & 1.52 & 6.30 & 1.96 & 0.36 & \nodata & 14.49 & 12.53 & 10.92 & 10.06 & 9.76 & 9.46 \\
670 & K4-M0 & 1.68 & 5.88 & 2.31 & 0.39 & -1.0 & 13.96 & 11.86 & 10.27 & 9.31 & 8.66 & 7.59 \\
672 & \nodata   & \nodata & \nodata & \nodata & \nodata & \nodata & 14.59 & 12.18 & 10.65 & 9.80 & 9.43 & 9.25 \\
718 & K4-M1 & 0.55 & 4.65 & 1.25 & 0.43 & \nodata & 13.83 & 11.55 & 10.08 & 9.17 & 8.73 & 8.45 \\
752 & M0 & 0.54 & 6.32 & 0.07 & 0.62 & 1.1 & 15.07 & 13.25 & 11.79 & 11.02 & 10.74 & \nodata \\
753 & K6 & 0.91 & 6.29 & 0.87 & 0.28 & 1.8 & 14.57 & 12.79 & 11.63 & 10.77 & 10.32 & \nodata \\
761 & K2-K4 & 1.35 & 6.70 & 2.55 & 1.46 & \nodata & 15.83 & 13.64 & 11.13 & 10.06 & 9.48 & 8.62 \\
801 & K4-M0 & 0.70 & 5.59 & 1.47 & 0.80 & -1.2 & 14.06 & 11.90 & 10.14 & 9.19 & 8.61 & 7.97 \\
828 & K2-K6 & 0.90 & 5.76 & 1.17 & 0.74 & 1.2 & 13.77 & 11.87 & 10.01 & 9.18 & 8.89 & 8.84 \\
848 & M2.5 & 0.29 & 6.07 & 1.72 & 0.50 & 0.0 & 17.52 & 14.47 & 12.43 & 11.67 & 11.30 & 10.59 \\
867 & K3-K7 & 2.62 & 5.56 & 2.76 & -0.17 & 1.6 & 13.04 & 10.82 & 9.44 & 8.56 & 8.19 & 7.95 \\
945 & M1.5 & 0.37 & 5.83 & 0.55 & 0.23 & 0.0 & 15.35 & 13.07 & 11.61 & 10.83 & 10.60 & \nodata \\
960 & M3.5 & 0.24 & 5.56 & 2.72 & -0.51 & 0.0 & 18.98 & 15.21 & 12.92 & 12.27 & 11.95 & \nodata \\
971 & K2.5-K7 & 0.69 & 5.77 & 0.00 & \nodata & 1.8 & 12.88 & 11.51 & 10.48 & 9.63 & 9.77 & \nodata \\
982 & K7 & 0.73 & 6.89 & 0.00 & \nodata & 1.8 & \nodata & 13.70 & 10.70 & 9.77 & 9.82 & \nodata \\
997 & \nodata   & \nodata & \nodata & \nodata & \nodata & \nodata & 15.63 & 13.42 & 11.63 & 10.68 & 10.31 & 9.92 \\
1002 & K2-K5 & 1.30 & 6.98 & 0.41 & 0.05 & 2.5 & 13.77 & 12.52 & 11.65 & 11.08 & 10.95 & \nodata \\
1083 & $<$M0 & \nodata & \nodata & \nodata & \nodata & \nodata & 15.15 & 12.96 & 11.12 & 10.08 & 9.72 & 9.24 \\
1111 & \nodata   & \nodata & \nodata & \nodata & \nodata & \nodata & 18.18 & 14.03 & 11.84 & 11.07 & 10.66 & 10.34 \\
1127 & K5.5-K7 & 0.90 & 6.05 & 3.69 & -0.18 & 1.4 & 16.93 & 14.05 & 12.08 & 11.03 & 10.64 & \nodata \\
1143 & K1-K2 & 1.90 & 6.54 & 2.16 & 0.24 & \nodata & 14.17 & 12.33 & 10.76 & 9.95 & 9.68 & 9.38 \\
1151 & K6 & 0.91 & 5.76 & 1.05 & 0.28 & 1.9 & 13.61 & 11.76 & 10.48 & 9.64 & 9.40 & \nodata \\
1246 & M3.5 & 0.23 & 6.26 & 0.92 & 0.75 & 0.0 & 17.80 & 14.73 & 12.60 & 11.56 & 10.96 & \nodata \\
1248 & M0.5 & 0.47 & 5.93 & 1.16 & 0.58 & 1.7 & 15.47 & 13.14 & 11.44 & 10.50 & 10.19 & \nodata \\
1252 & M0 & 0.53 & 6.27 & 1.22 & 0.65 & 1.6 & 16.14 & 13.87 & 11.99 & 11.12 & 10.79 & \nodata \\
1261 & \nodata   & \nodata & \nodata & \nodata & \nodata & \nodata & \nodata & 13.20 & 12.41 & 11.63 & 11.27 & 10.93 \\
1269 & G8-K3 & 2.01 & 6.51 & 0.75 & 0.10 & \nodata & 12.34 & 11.12 & 10.13 & 9.54 & 9.41 & \nodata \\
1311 & K2-K4 & 1.53 & 6.23 & 1.80 & 0.06 & \nodata & 14.23 & 12.33 & 11.01 & 10.17 & 9.94 & 9.92 \\
1350 & G3-K3 & 2.20 & 6.60 & 1.15 & 0.12 & 1.7 & 11.78 & 10.59 & 9.70 & 9.15 & 8.97 & \nodata \\
1355 & M3.5 & 0.24 & 6.07 & 0.00 & 0.26 & 0.0 & 16.30 & 13.96 & 12.32 & 11.63 & 11.37 & \nodata \\
1374 & \nodata   & \nodata & \nodata & \nodata & \nodata & \nodata & \nodata & 15.22 & 12.75 & 11.76 & 11.16 & \nodata \\
1384 & K5-M0.5e & 0.52 & 5.95 & 0.00 & 0.56 & 1.9 & 14.12 & 12.37 & 10.95 & 10.18 & 9.96 & \nodata \\
1412 & M1.5-M4 & 0.37 & 5.66 & 0.75 & 0.63 & 1.8 & 15.36 & 13.00 & 11.56 & 10.65 & 10.39 & 9.88 \\
1424 & M0-M1 & 0.53 & 6.06 & 1.07 & 0.62 & 1.2 & 15.51 & 13.30 & 11.53 & 10.67 & 10.35 & \nodata \\
1429 & M1 & 0.43 & 4.14 & 4.42 & -1.41 & 1.2 & 17.20 & 13.50 & 12.00 & 11.11 & 10.90 & 10.60 \\
1433 & \nodata   & \nodata & \nodata & \nodata & \nodata & \nodata & 17.29 & 13.89 & 12.02 & 11.22 & 10.93 & \nodata \\
1443 & \nodata   & \nodata & \nodata & \nodata & \nodata & \nodata & 14.37 & 12.48 & 11.11 & 10.31 & 10.08 & \nodata \\
1449 & \nodata   & \nodata & \nodata & \nodata & \nodata & \nodata & 17.63 & 14.77 & 12.42 & 11.35 & 10.91 & \nodata \\
1463 & K8e-M1: & 0.59 & 5.83 & 0.56 & 0.50 & \nodata & 14.12 & 12.19 & 10.78 & 10.00 & 9.50 & \nodata \\
1487 & M1 (SB2) & 0.40 & 5.86 & 1.65 & \nodata & \nodata & 15.88 & 13.26 & 11.55 & 10.61 & 10.28 & \nodata \\
1489 & F9-K0 & 2.59 & 6.46 & 2.06 & 0.05 & \nodata & 11.72 & 10.30 & 9.31 & 8.61 & 8.40 & \nodata \\
1492 & M1.5 & 0.37 & 6.03 & 0.11 & 0.46 & 1.6 & 15.13 & 13.02 & 11.63 & 10.88 & 10.57 & \nodata \\
1499 & \nodata   & \nodata & \nodata & \nodata & \nodata & \nodata & 15.84 & 13.79 & 11.39 & 10.46 & 9.85 & \nodata \\
1516 & K1-K4 & 1.40 & 6.82 & 1.31 & -0.20 & 1.8 & 14.37 & 12.77 & 11.72 & 11.05 & 10.89 & \nodata \\
1521 & K4 & 1.40 & 6.67 & 1.08 & 1.01 & \nodata & 14.31 & 12.69 & 11.24 & 10.36 & 9.83 & \nodata \\
1568 & K0-K1e & 2.55 & 6.34 & 0.59 & 0.19 & \nodata & 11.30 & 10.20 & 9.36 & 8.84 & 8.63 & \nodata \\
1595 & M2.5 & 0.29 & 5.99 & 0.00 & 0.11 & \nodata & 15.57 & 13.25 & 11.89 & 11.19 & 10.97 & \nodata \\
1608 & M0.5e & 0.48 & 6.23 & 0.93 & 1.41 & -1.3 & 16.01 & 13.77 & 11.96 & 11.06 & 10.41 & \nodata \\
\cutinhead{High-absorption sample} \\
90 & M0 & 0.52 & 5.93 & 4.97 & 0.08 & 1.6 & 19.09 & 15.36 & 12.68 & 11.44 & 10.97 & \nodata \\
115 & K7 & 0.71 & 6.21 & 3.83 & 0.86 & 1.4 & 17.99 & 14.91 & 12.20 & 10.91 & 10.43 & \nodata \\
131 & K5 & 1.20 & 6.34 & 3.95 & 0.53 & 1.4 & 17.12 & 14.27 & 11.98 & 10.95 & 10.24 & \nodata \\
183 & G: & \nodata & \nodata & \nodata & \nodata & 0.0 & 18.03 & 15.68 & 12.03 & 10.31 & 9.23 & \nodata \\
223 & K5 & 1.19 & 6.08 & 4.66 & 1.04 & 1.7 & 17.35 & 14.22 & 11.53 & 10.10 & 9.34 & \nodata \\
262 & K5 & 1.13 & 6.78 & 3.77 & 2.24 & 2.3 & 17.69 & 14.91 & 11.66 & 10.07 & 9.30 & 8.59 \\
310 & \nodata   & \nodata & \nodata & \nodata & \nodata & 8.9 & 17.91 & 14.65 & 11.76 & 10.40 & 9.62 & \nodata \\
323 & K6-M0 & 0.57 & 6.21 & 3.86 & 1.22 & 2.0 & 18.50 & 15.24 & 12.44 & 11.11 & 10.46 & \nodata \\
331 & \nodata   & \nodata & \nodata & \nodata & \nodata & \nodata & \nodata & 16.23 & 12.48 & 10.55 & 9.36 & \nodata \\
342 & \nodata   & \nodata & \nodata & \nodata & \nodata & \nodata & \nodata & 14.89 & 11.73 & 10.25 & 9.62 & 9.04 \\
365 & K4-K7 & 0.72 & 7.79 & 0.00 & \nodata & 0.0 & \nodata & 14.37 & 11.14 & 9.66 & 8.82 & 7.66 \\
449 & K7 & \nodata & \nodata & \nodata & \nodata & 0.0 & 17.00 & 15.38 & 12.26 & 10.45 & 9.38 & 8.05 \\
452 & K0-K3 & 2.02 & 6.17 & 5.36 & 0.61 & 1.4 & 17.04 & 13.86 & 11.03 & 9.73 & 8.99 & \nodata \\
454 & K2-K7 & 2.35 & 5.91 & 5.85 & 0.20 & 2.1 & 16.85 & 13.48 & 10.83 & 9.61 & 9.10 & 8.21 \\
490 & K6-K8 & 0.70 & 5.62 & 4.91 & 0.32 & 1.2 & 17.55 & 14.05 & 11.40 & 10.10 & 9.61 & \nodata \\
499 & K5-M1 & 0.69 & 5.92 & 2.65 & 0.06 & 1.2 & 16.19 & 13.57 & 11.68 & 10.71 & 10.35 & \nodata \\
514 & \nodata   & \nodata & \nodata & \nodata & \nodata & \nodata & 18.20 & 15.10 & 12.36 & 10.93 & 10.37 & \nodata \\
554 & \nodata   & \nodata & \nodata & \nodata & \nodata & \nodata & \nodata & \nodata & 12.50 & 12.72 & 10.42 & 8.25 \\
561 & K5 & \nodata & \nodata & 0.00 & \nodata & 1.0 & \nodata & 14.58 & 11.12 & 9.41 & 8.34 & 6.67 \\
626 & M1 & 0.41 & 6.02 & 3.47 & 0.62 & 0.7 & 18.30 & 14.97 & 12.38 & 11.19 & 10.78 & \nodata \\
649 & M0.5-M2.5 & 0.40 & 6.03 & 4.11 & 0.43 & 0.0 & 19.12 & 15.45 & 12.64 & 11.38 & 10.84 & 10.46 \\
655 & \nodata   & \nodata & \nodata & \nodata & \nodata & \nodata & \nodata & \nodata & \nodata & 13.40 & 10.38 & 7.70 \\
682 & K2-K5 & 1.86 & 6.05 & 2.56 & \nodata & \nodata & 14.26 & 12.12 & 10.78 & 9.26 & 8.63 & 7.26 \\
697 & K5-lateK & \nodata & \nodata & \nodata & \nodata & 6.1 & 15.17 & 12.54 & 10.25 & 9.11 & 8.06 & 6.84 \\
707 & M2 & 0.33 & 5.38 & 1.21 & 0.83 & 1.6 & 16.04 & 13.34 & 11.41 & 10.36 & 9.77 & 8.75 \\
720 & \nodata   & \nodata & \nodata & \nodata & \nodata & \nodata & \nodata & 16.57 & 12.72 & 10.93 & 10.03 & \nodata \\
758 & G5-K0e & 3.00 & 6.12 & 3.78 & 1.50 & -12.3 & 13.79 & 11.45 & 8.78 & 7.76 & 7.13 & 6.16 \\
766 & \nodata   & \nodata & \nodata & \nodata & \nodata & \nodata & \nodata & 11.00 & 9.74 & 8.37 & 7.35 & 5.50 \\
784 & M1.5-M2e & \nodata & \nodata & 0.00 & \nodata & \nodata & \nodata & 17.99 & 12.20 & 11.20 & 10.70 & 8.63 \\
874 & \nodata   & \nodata & \nodata & \nodata & \nodata & \nodata & \nodata & \nodata & \nodata & 14.31 & 12.46 & 10.36 \\
894 & \nodata   & \nodata & \nodata & \nodata & \nodata & \nodata & \nodata & \nodata & 13.81 & 11.82 & 10.95 & \nodata \\
915 & \nodata   & \nodata & \nodata & \nodata & \nodata & \nodata & \nodata & \nodata & 13.63 & 11.21 & 9.98 & \nodata \\
939 & \nodata   & \nodata & \nodata & \nodata & \nodata & \nodata & 18.67 & 14.93 & 11.49 & 9.97 & 8.86 & \nodata \\
942 & M0 & 0.52 & 5.90 & 5.04 & \nodata & 0.0 & 19.06 & 15.30 & 11.98 & 10.43 & 9.67 & 9.14 \\
985 & F8-K0 & 2.97 & 6.17 & 2.06 & 1.19 & \nodata & 12.23 & 10.56 & 8.69 & 7.75 & 7.37 & 6.91 \\
1028 & K2 & 1.60 & 6.78 & 2.72 & 1.45 & \nodata & 15.26 & 13.20 & 10.93 & 9.75 & 9.11 & 8.53 \\
1035 & \nodata   & \nodata & \nodata & \nodata & \nodata & \nodata & \nodata & \nodata & 13.62 & 11.25 & 10.03 & \nodata \\
1040 & \nodata   & \nodata & \nodata & \nodata & \nodata & \nodata & \nodata & \nodata & 13.99 & 11.49 & 10.16 & \nodata \\
1071 & K7-M0 & 0.69 & 5.92 & 1.57 & 1.29 & 1.6 & 15.09 & 12.89 & 10.38 & 9.26 & 8.38 & 7.38 \\
1080 & earlyKe-M0 & 1.98 & 5.25 & 7.77 & 1.41 & -16.9 & 17.82 & 13.48 & 9.65 & 7.72 & 6.43 & \nodata \\
1114 & K0-K5: & \nodata & \nodata & \nodata & \nodata & -1.5 & 15.72 & 12.46 & 9.80 & 8.58 & 8.04 & \nodata \\
1140 & \nodata   & \nodata & \nodata & \nodata & \nodata & \nodata & 19.59 & 15.89 & 12.80 & 11.28 & 10.40 & 9.57 \\
1158 & M1: & 0.41 & 6.08 & 2.65 & \nodata & 0.0 & 17.67 & 14.66 & 11.90 & 10.54 & 9.70 & 8.51 \\
1161 & M0 & 0.54 & 6.43 & 0.00 & 1.12 & 1.5 & 15.21 & 13.44 & 11.80 & 10.90 & 10.51 & 10.11 \\
1304 & \nodata   & \nodata & \nodata & \nodata & \nodata & \nodata & \nodata & \nodata & \nodata & \nodata & \nodata & \nodata \\
1309 & K1-K4 & 1.90 & 6.44 & 4.33 & \nodata & \nodata & 16.37 & 13.64 & 10.64 & 9.39 & 8.75 & 7.87 \\
1335 & K8e & 0.64 & 6.64 & 1.10 & 2.18 & -2.5 & 16.32 & 14.18 & 11.97 & 10.74 & 9.99 & 8.65 \\
1341 & \nodata   & \nodata & \nodata & \nodata & \nodata & \nodata & \nodata & 14.42 & 11.42 & 9.96 & 9.32 & 8.78 \\
1343 & \nodata   & \nodata & \nodata & \nodata & \nodata & \nodata & 17.65 & 14.63 & 12.02 & 10.61 & 9.78 & 8.83 \\
1354 & G0-M3 & 2.34 & 6.55 & 1.70 & 0.18 & \nodata & 11.97 & 10.61 & 9.67 & 9.04 & 8.80 & 8.25 \\
1380 & K4 & 1.52 & 6.27 & 3.78 & 0.87 & \nodata & 16.28 & 13.61 & 11.69 & 10.31 & 9.27 & 8.25 \\
1382 & M0 & 0.52 & 5.88 & 1.97 & 1.25 & 0.0 & 15.94 & 13.38 & 11.31 & 10.17 & 9.45 & \nodata \\
1391 & M1 & 0.41 & 6.03 & 3.80 & 1.53 & 1.4 & 18.69 & 15.23 & 12.14 & 10.57 & 9.92 & \nodata \\
1409 & K6-K8e & 0.74 & 7.10 & 0.00 & 3.01 & -6.3 & 15.36 & 13.98 & 11.70 & 10.15 & 9.20 & 8.07 \\
1410 & M1 & 0.36 & 7.56 & 0.57 & 2.30 & 0.0 & 18.54 & 16.34 & 13.60 & 12.31 & 11.85 & \nodata \\
1421 & M0 & 0.54 & 6.13 & 1.04 & 1.10 & 0.9 & 15.63 & 13.43 & 11.60 & 10.71 & 10.33 & \nodata \\
1444 & K8e & 0.60 & 6.19 & 1.10 & 0.73 & -4.1 & 15.55 & 13.41 & 11.78 & 10.94 & 10.43 & \nodata \\
1456 & \nodata   & \nodata & \nodata & \nodata & \nodata & \nodata & \nodata & 16.83 & 13.36 & 11.44 & 10.57 & \nodata \\
1462 & K1 & 2.54 & 6.25 & 3.09 & 0.18 & \nodata & 13.95 & 11.82 & 10.12 & 9.25 & 8.92 & \nodata \\
1466 & K3-K5 & 1.40 & 6.65 & 1.96 & 0.42 & 1.4 & 15.15 & 13.19 & 11.69 & 10.78 & 10.38 & \nodata \\
\enddata
\normalsize
\end{deluxetable}

%% file: tab2.tex
\begin{deluxetable}{rcccrrrrccccccccccrr}
\tablecaption{Spectral analysis results
\label{tab:res}}
\tabletypesize{\scriptsize}
\tablehead{
\multicolumn{1}{c}{COUP} & $N_{\rm H}$ & $T_1$ & $T_2$ & $T_3$ & $k_1$ & $k_2$ & $k_3$ &
\multicolumn{9}{c}{Abundances (solar units)} & & &
\multicolumn{1}{c}{$f_{\rm x}$} \\
\multicolumn{1}{c}{ID} & (a) & keV & keV & keV & (b) & (b) & (b) & O & Ne & Mg & Si & S &
Ar & Ca
& Fe & Ni & $\chi^2_{\rm r}$ & DoF & \multicolumn{1}{c}{(c)}
}
\startdata
7 & 1.5 & 0.2 & 0.7 & 2.2 & 0.6 & 2.2 & 3.8 & 0.4 & 1.0 & 0.3 & 0.2 & 0.3 & 0.7 & 0.2 & 0.2 & 0.7 & 1.2 & 162 & 1.5 \\
9 & 2.6 & \nodata & 0.8 & 3.1 & \nodata & 1.6 & 2.8 & 0.9 & 0.6 & 0.4 & 0.4 & 0.6 & 1.3 & 0.0 & 0.3 & 1.7 & 1.3 & 125 & 1.7 \\
11 & 3.9 & \nodata & 0.6 & 5.4 & \nodata & 0.9 & 0.7 & 0.1 & 0.3 & 0.2 & 0.7 & 2.9 & 2.5 & 0.0 & 0.1 & 0.7 & 1.6 & 61 & 0.6 \\
23 & 2.4 & 0.1 & 0.8 & 2.4 & 0.8 & 3.6 & 5.6 & 0.8 & 0.9 & 0.2 & 0.1 & 0.5 & 0.6 & 0.1 & 0.2 & 0.7 & 1.2 & 203 & 2.6 \\
27 & 1.7 & \nodata & 0.7 & 2.7 & \nodata & 0.3 & 0.5 & 0.7 & 1.4 & 0.2 & 0.2 & 0.6 & 1.3 & 0.0 & 0.2 & 0.4 & 0.9 & 61 & 0.3 \\
28 & 1.5 & \nodata & 0.6 & 3.4 & \nodata & 0.5 & 2.5 & 0.6 & 1.1 & 0.2 & 0.3 & 0.9 & 0.5 & 0.0 & 0.2 & 0.7 & 1.2 & 154 & 1.6 \\
43 & 5.3 & \nodata & 0.3 & 2.6 & \nodata & 18.6 & 1.0 & 0.0 & 0.0 & 0.0 & 0.1 & 0.6 & 0.0 & 0.0 & 0.0 & 1.6 & 1.1 & 74 & 0.4 \\
62 & 2.6 & \nodata & 0.6 & 3.1 & \nodata & 1.0 & 0.9 & 0.4 & 0.2 & 0.1 & 0.2 & 1.0 & 1.6 & 0.0 & 0.0 & 1.8 & 1.2 & 97 & 0.5 \\
66 & 1.7 & \nodata & 0.7 & 3.2 & \nodata & 0.2 & 0.6 & 1.2 & 1.3 & 0.0 & 0.0 & 1.5 & 0.6 & 0.0 & 0.1 & 0.6 & 1.2 & 64 & 0.4 \\
67 & 2.0 & \nodata & 0.7 & 2.4 & \nodata & 0.4 & 0.7 & 0.9 & 1.1 & 0.5 & 0.2 & 0.3 & 1.4 & 0.0 & 0.3 & 0.8 & 1.0 & 75 & 0.3 \\
71 & 0.0 & \nodata & 0.7 & 2.9 & \nodata & 0.6 & 0.4 & 0.3 & 0.1 & 0.0 & 0.0 & 0.9 & 0.9 & 1.8 & 0.1 & 1.3 & 1.0 & 62 & 0.2 \\
101 & 1.6 & \nodata & 0.9 & 3.7 & \nodata & 0.4 & 1.3 & 0.9 & 0.3 & 0.3 & 0.0 & 0.8 & 1.4 & 1.4 & 0.1 & 2.8 & 0.8 & 103 & 0.9 \\
108 & 3.1 & \nodata & 0.4 & 2.9 & \nodata & 0.2 & 1.0 & 0.7 & 2.1 & 0.9 & 0.5 & 0.0 & 0.6 & 0.1 & 0.3 & 2.9 & 0.9 & 69 & 0.6 \\
112 & 3.2 & \nodata & 0.6 & 2.5 & \nodata & 0.2 & 0.7 & 1.2 & 2.1 & 0.4 & 0.5 & 0.6 & 0.9 & 0.9 & 0.6 & 1.6 & 1.1 & 77 & 0.4 \\
113 & 2.9 & \nodata & 0.8 & 2.3 & \nodata & 1.1 & 2.0 & 1.1 & 0.9 & 0.1 & 0.1 & 0.8 & 0.6 & 0.0 & 0.2 & 1.3 & 1.3 & 71 & 0.9 \\
139 & 1.6 & \nodata & 0.8 & 2.7 & \nodata & 0.6 & 0.4 & 0.5 & 0.2 & 0.0 & 0.0 & 0.4 & 1.5 & 0.0 & 0.1 & 0.0 & 0.9 & 62 & 0.2 \\
141 & 1.8 & \nodata & 0.8 & 4.0 & \nodata & 1.0 & 1.0 & 0.8 & 0.5 & 0.1 & 0.2 & 1.0 & 1.1 & 0.0 & 0.2 & 0.1 & 1.1 & 117 & 0.8 \\
150 & 2.8 & \nodata & 0.7 & 2.8 & \nodata & 0.3 & 0.6 & 0.9 & 1.1 & 0.2 & 0.2 & 0.9 & 1.7 & 0.3 & 0.2 & 1.1 & 1.3 & 64 & 0.3 \\
152 & 0.5 & \nodata & 0.7 & 2.7 & \nodata & 0.3 & 0.4 & 0.7 & 0.4 & 0.0 & 0.1 & 0.6 & 0.0 & 1.7 & 0.1 & 1.5 & 1.2 & 58 & 0.2 \\
173 & 1.9 & \nodata & 0.8 & 2.7 & \nodata & 1.0 & 0.8 & 1.0 & 0.1 & 0.0 & 0.1 & 0.7 & 0.9 & 0.3 & 0.2 & 0.2 & 1.0 & 99 & 0.5 \\
177 & 4.6 & \nodata & 0.8 & 3.1 & \nodata & 1.2 & 0.4 & 0.0 & 0.0 & 0.1 & 0.1 & 0.3 & 1.1 & 0.1 & 0.1 & 0.1 & 0.8 & 48 & 0.3 \\
188 & 2.7 & \nodata & 0.7 & 2.5 & \nodata & 1.8 & 3.4 & 0.6 & 0.9 & 0.1 & 0.1 & 0.4 & 0.5 & 0.0 & 0.1 & 1.1 & 1.3 & 158 & 1.5 \\
202 & 2.4 & \nodata & 0.5 & 2.2 & \nodata & 0.3 & 0.6 & 0.3 & 0.7 & 0.0 & 0.2 & 0.4 & 0.9 & 0.0 & 0.1 & 2.1 & 1.3 & 49 & 0.2 \\
205 & 1.9 & \nodata & 0.8 & 4.0 & \nodata & 1.1 & 0.5 & 0.3 & 0.2 & 0.0 & 0.1 & 1.0 & 2.7 & 0.8 & 0.1 & 0.0 & 1.2 & 66 & 0.4 \\
270 & 1.3 & \nodata & 0.8 & 2.3 & \nodata & 0.3 & 0.4 & 1.4 & 0.8 & 0.2 & 0.4 & 0.6 & 0.8 & 0.5 & 0.3 & 0.0 & 1.2 & 63 & 0.2 \\
328 & 1.5 & \nodata & 0.6 & 2.0 & \nodata & 1.2 & 1.0 & 0.5 & 0.8 & 0.2 & 0.1 & 0.4 & 0.5 & 0.0 & 0.1 & 1.7 & 1.2 & 97 & 0.4 \\
343 & 1.6 & \nodata & 0.7 & 3.1 & \nodata & 1.1 & 2.8 & 1.2 & 1.7 & 0.1 & 0.3 & 1.2 & 0.8 & 0.4 & 0.2 & 0.1 & 1.6 & 177 & 1.8 \\
382 & 2.7 & \nodata & 0.6 & 2.3 & \nodata & 0.2 & 0.3 & 1.9 & 2.3 & 0.7 & 1.0 & 1.7 & 0.2 & 4.9 & 0.4 & 3.9 & 0.9 & 50 & 0.2 \\
387 & 2.4 & \nodata & 0.7 & 2.5 & \nodata & 0.9 & 2.1 & 0.5 & 1.1 & 0.2 & 0.2 & 0.7 & 0.4 & 0.3 & 0.2 & 1.0 & 1.2 & 135 & 1.0 \\
417 & 1.9 & \nodata & 0.8 & 3.0 & \nodata & 0.5 & 0.6 & 0.5 & 0.7 & 0.3 & 0.1 & 1.1 & 1.2 & 1.4 & 0.1 & 1.2 & 0.8 & 62 & 0.4 \\
431 & 5.1 & \nodata & 0.9 & 2.4 & \nodata & 1.0 & 2.7 & 0.9 & 1.4 & 0.3 & 0.4 & 0.6 & 0.6 & 0.8 & 0.3 & 2.8 & 1.1 & 141 & 1.3 \\
459 & 0.2 & \nodata & 0.7 & 2.7 & \nodata & 0.1 & 0.6 & 0.7 & 2.4 & 0.8 & 0.7 & 0.4 & 1.6 & 0.0 & 0.4 & 1.9 & 0.9 & 78 & 0.3 \\
470 & 1.4 & \nodata & 0.8 & 2.7 & \nodata & 0.5 & 0.9 & 1.5 & 0.6 & 0.2 & 0.2 & 1.1 & 1.1 & 0.0 & 0.4 & 2.2 & 1.3 & 99 & 0.5 \\
567 & 2.1 & \nodata & 0.7 & 3.0 & \nodata & 0.5 & 1.0 & 0.7 & 1.3 & 0.5 & 0.4 & 0.8 & 1.4 & 0.5 & 0.2 & 1.5 & 1.0 & 99 & 0.6 \\
579 & 4.5 & \nodata & 0.4 & 4.4 & \nodata & 0.1 & 1.0 & 0.1 & 2.5 & 0.0 & 1.1 & 1.6 & 3.8 & 0.0 & 0.2 & 0.0 & 1.6 & 79 & 0.7 \\
597 & 0.0 & \nodata & 0.7 & 2.4 & \nodata & 0.2 & 0.4 & 1.0 & 1.6 & 0.0 & 0.5 & 1.1 & 1.7 & 5.7 & 0.2 & 3.3 & 1.0 & 78 & 0.3 \\
600 & 2.7 & \nodata & 0.6 & 2.4 & \nodata & 0.2 & 0.4 & 0.8 & 1.6 & 0.8 & 0.9 & 1.7 & 0.3 & 0.0 & 0.3 & 3.1 & 0.9 & 52 & 0.2 \\
648 & 3.8 & \nodata & 0.6 & 2.7 & \nodata & 0.7 & 2.8 & 1.0 & 2.3 & 0.4 & 0.5 & 1.2 & 0.9 & 0.0 & 0.2 & 1.7 & 1.8 & 159 & 1.6 \\
669 & 3.6 & \nodata & 0.6 & 3.0 & \nodata & 0.5 & 2.4 & 1.0 & 2.3 & 0.5 & 0.5 & 0.8 & 0.9 & 0.0 & 0.3 & 0.6 & 1.2 & 149 & 1.4 \\
670 & 4.4 & 0.5 & 2.2 & 7.7 & 1.0 & 2.9 & 1.0 & 0.6 & 2.0 & 0.6 & 0.6 & 1.0 & 1.7 & 0.0 & 0.2 & 1.2 & 1.6 & 185 & 2.3 \\
672 & 2.6 & \nodata & 0.4 & 2.5 & \nodata & 0.5 & 1.2 & 0.3 & 1.4 & 0.7 & 0.8 & 1.3 & 0.5 & 0.0 & 0.2 & 1.0 & 1.5 & 80 & 0.6 \\
718 & 3.8 & \nodata & 0.6 & 2.9 & \nodata & 0.7 & 2.5 & 0.0 & 1.2 & 0.2 & 0.3 & 0.8 & 0.0 & 0.0 & 0.2 & 1.8 & 1.2 & 143 & 1.2 \\
752 & 2.8 & 0.4 & 1.3 & 4.2 & 1.0 & 1.2 & 3.8 & 0.3 & 0.9 & 0.2 & 0.1 & 0.3 & 0.8 & 0.0 & 0.2 & 1.5 & 0.7 & 199 & 2.8 \\
753 & 5.3 & \nodata & 0.4 & 2.1 & \nodata & 0.7 & 1.0 & 0.5 & 0.9 & 0.5 & 0.3 & 0.5 & 1.0 & 0.0 & 0.2 & 0.0 & 1.0 & 63 & 0.4 \\
761 & 5.5 & \nodata & 0.8 & 2.8 & \nodata & 0.3 & 0.8 & 7.3 & 2.1 & 0.5 & 0.3 & 0.8 & 1.0 & 0.6 & 0.9 & 1.2 & 1.1 & 108 & 0.7 \\
801 & 5.9 & \nodata & 0.4 & 3.6 & \nodata & 0.6 & 3.6 & 4.8 & 4.6 & 1.5 & 1.3 & 2.0 & 2.9 & 0.0 & 0.5 & 2.4 & 1.3 & 127 & 3.4 \\
828 & 3.5 & \nodata & 0.5 & 4.0 & \nodata & 0.3 & 4.2 & 0.3 & 4.4 & 0.4 & 1.2 & 1.6 & 1.4 & 0.0 & 0.4 & 1.4 & 1.1 & 123 & 3.2 \\
848 & 2.7 & 0.5 & 1.7 & 10.8 & 0.3 & 0.6 & 0.3 & 0.5 & 0.7 & 0.0 & 0.3 & 0.9 & 1.3 & 0.0 & 0.1 & 1.2 & 1.2 & 77 & 0.5 \\
867 & 4.1 & \nodata & 0.6 & 2.7 & \nodata & 0.4 & 1.9 & 0.3 & 2.3 & 0.6 & 0.7 & 0.6 & 0.5 & 0.0 & 0.4 & 3.7 & 1.2 & 128 & 1.0 \\
945 & 1.0 & \nodata & 0.4 & 1.9 & \nodata & 0.3 & 0.5 & 0.4 & 1.4 & 0.7 & 0.4 & 0.9 & 0.0 & 3.0 & 0.2 & 1.3 & 0.9 & 63 & 0.2 \\
960 & 1.1 & \nodata & 0.7 & 3.3 & \nodata & 0.0 & 0.4 & 1.4 & 2.3 & 0.4 & 0.4 & 1.2 & 2.5 & 0.0 & 1.0 & 3.2 & 1.2 & 56 & 0.3 \\
971 & 0.3 & 0.2 & 0.8 & 2.6 & 0.2 & 1.2 & 3.6 & 0.6 & 1.6 & 0.0 & 0.4 & 0.5 & 0.9 & 0.0 & 0.2 & 1.9 & 1.3 & 142 & 1.8 \\
982 & 2.7 & \nodata & 1.3 & 4.7 & \nodata & 1.2 & 1.2 & 0.0 & 1.4 & 0.0 & 0.5 & 0.7 & 0.4 & 3.2 & 0.0 & 0.3 & 1.1 & 63 & 1.1 \\
997 & 3.6 & \nodata & 0.7 & 3.0 & \nodata & 0.7 & 2.0 & 0.0 & 1.0 & 0.3 & 0.3 & 1.3 & 0.2 & 0.0 & 0.1 & 1.5 & 1.0 & 130 & 1.1 \\
1002 & 0.6 & \nodata & 0.4 & 1.9 & \nodata & 0.9 & 0.7 & 0.2 & 0.8 & 0.1 & 0.2 & 0.5 & 0.1 & 0.3 & 0.1 & 1.3 & 0.9 & 76 & 0.2 \\
1083 & 5.0 & \nodata & 0.6 & 2.6 & \nodata & 0.7 & 2.1 & 0.3 & 1.5 & 0.1 & 0.2 & 0.7 & 0.8 & 0.0 & 0.2 & 1.8 & 1.2 & 124 & 1.0 \\
1111 & 3.8 & \nodata & 0.6 & 2.8 & \nodata & 0.2 & 0.8 & 2.6 & 2.4 & 0.7 & 0.5 & 1.0 & 0.7 & 0.0 & 0.4 & 2.1 & 1.3 & 81 & 0.5 \\
1127 & 5.4 & \nodata & 0.5 & 2.0 & \nodata & 0.4 & 1.1 & 0.0 & 1.3 & 0.4 & 0.5 & 0.5 & 1.5 & 0.0 & 0.1 & 0.1 & 1.2 & 60 & 0.4 \\
1143 & 1.8 & \nodata & 0.4 & 2.2 & \nodata & 0.4 & 1.7 & 0.3 & 1.8 & 0.5 & 0.4 & 0.6 & 1.0 & 0.0 & 0.1 & 1.2 & 1.2 & 119 & 0.7 \\
1151 & 1.6 & \nodata & 0.7 & 2.6 & \nodata & 1.5 & 2.1 & 0.5 & 0.8 & 0.0 & 0.1 & 0.9 & 0.9 & 0.0 & 0.1 & 1.3 & 1.4 & 146 & 1.0 \\
1246 & 4.3 & \nodata & 0.4 & 3.4 & \nodata & 0.3 & 1.0 & 1.3 & 1.8 & 0.2 & 0.5 & 0.1 & 0.3 & 0.0 & 0.3 & 1.3 & 1.2 & 85 & 0.6 \\
1248 & 3.0 & \nodata & 0.8 & 2.5 & \nodata & 0.8 & 1.8 & 0.0 & 0.7 & 0.1 & 0.2 & 0.7 & 0.7 & 0.0 & 0.1 & 1.7 & 1.1 & 117 & 0.8 \\
1252 & 4.7 & \nodata & 0.7 & 3.3 & \nodata & 0.3 & 0.6 & 1.4 & 1.5 & 0.6 & 0.6 & 1.1 & 0.6 & 0.1 & 0.5 & 1.7 & 1.1 & 59 & 0.4 \\
1261 & 2.8 & \nodata & 0.6 & 3.5 & \nodata & 0.1 & 0.5 & 2.2 & 1.9 & 0.5 & 0.6 & 1.2 & 0.0 & 0.0 & 0.4 & 2.7 & 1.4 & 57 & 0.4 \\
1269 & 1.5 & 0.4 & 0.8 & 1.8 & 1.1 & 1.4 & 2.7 & 0.4 & 1.1 & 0.4 & 0.1 & 0.3 & 0.6 & 0.0 & 0.2 & 0.3 & 1.1 & 136 & 0.9 \\
1311 & 2.1 & \nodata & 0.5 & 2.2 & \nodata & 0.2 & 0.4 & 0.5 & 2.0 & 0.4 & 0.7 & 1.4 & 0.7 & 1.6 & 0.3 & 3.2 & 1.0 & 51 & 0.2 \\
1350 & 2.2 & 0.5 & 0.9 & 2.3 & 1.0 & 1.2 & 2.8 & 0.6 & 0.8 & 0.3 & 0.3 & 0.6 & 0.5 & 0.1 & 0.2 & 1.2 & 1.4 & 145 & 1.3 \\
1355 & 1.5 & \nodata & 0.4 & 2.0 & \nodata & 0.5 & 0.6 & 0.4 & 0.8 & 0.1 & 0.3 & 0.4 & 0.0 & 0.0 & 0.1 & 1.1 & 1.2 & 58 & 0.2 \\
1374 & 4.2 & \nodata & 0.4 & 2.6 & \nodata & 0.1 & 0.9 & 0.0 & 2.1 & 0.3 & 0.2 & 0.8 & 0.7 & 1.0 & 0.2 & 1.3 & 1.3 & 56 & 0.4 \\
1384 & 1.0 & \nodata & 0.4 & 3.2 & \nodata & 0.2 & 2.4 & 0.7 & 2.7 & 0.1 & 0.6 & 0.4 & 0.5 & 0.8 & 0.2 & 2.3 & 1.1 & 160 & 1.5 \\
1412 & 1.6 & \nodata & 0.8 & 2.2 & \nodata & 0.3 & 0.6 & 0.7 & 0.7 & 0.4 & 0.3 & 1.1 & 0.3 & 1.8 & 0.3 & 1.2 & 1.0 & 74 & 0.3 \\
1424 & 3.1 & \nodata & 0.5 & 2.1 & \nodata & 0.4 & 0.9 & 0.3 & 1.5 & 0.2 & 0.2 & 0.4 & 0.2 & 0.0 & 0.1 & 2.2 & 1.1 & 74 & 0.3 \\
1429 & 1.1 & \nodata & 0.4 & 2.0 & \nodata & 0.2 & 0.6 & 0.3 & 2.0 & 0.0 & 0.3 & 0.4 & 0.1 & 0.0 & 0.2 & 0.6 & 1.1 & 52 & 0.2 \\
1433 & 4.2 & \nodata & 0.4 & 2.5 & \nodata & 0.2 & 0.6 & 3.7 & 4.1 & 0.0 & 0.3 & 0.4 & 0.2 & 0.0 & 0.5 & 5.0 & 0.9 & 71 & 0.4 \\
1443 & 2.2 & \nodata & 0.7 & 3.4 & \nodata & 0.2 & 0.8 & 1.7 & 2.0 & 0.4 & 0.8 & 1.3 & 0.0 & 0.4 & 0.5 & 2.3 & 1.2 & 97 & 0.6 \\
1449 & 4.0 & \nodata & 0.4 & 3.3 & \nodata & 0.0 & 0.8 & 0.0 & 4.4 & 0.3 & 1.0 & 1.5 & 0.8 & 0.0 & 0.4 & 4.7 & 1.2 & 61 & 0.5 \\
1463 & 0.7 & \nodata & 0.4 & 2.5 & \nodata & 0.2 & 0.8 & 0.4 & 1.7 & 0.2 & 0.4 & 0.4 & 0.2 & 0.0 & 0.3 & 1.9 & 1.1 & 79 & 0.3 \\
1487 & 4.2 & \nodata & 0.4 & 2.2 & \nodata & 0.3 & 0.6 & 1.0 & 2.2 & 0.8 & 0.7 & 0.8 & 0.8 & 0.7 & 0.4 & 4.6 & 1.2 & 60 & 0.3 \\
1489 & 1.9 & 0.4 & 0.8 & 2.0 & 0.4 & 0.5 & 1.5 & 0.4 & 1.2 & 0.5 & 0.4 & 0.7 & 0.3 & 0.3 & 0.2 & 1.2 & 1.4 & 107 & 0.5 \\
1492 & 1.3 & \nodata & 0.7 & 2.4 & \nodata & 0.2 & 0.5 & 1.2 & 2.1 & 0.7 & 0.6 & 0.9 & 0.1 & 0.0 & 0.3 & 2.4 & 1.0 & 67 & 0.2 \\
1499 & 5.9 & \nodata & 0.6 & 4.0 & \nodata & 0.2 & 0.7 & 4.0 & 2.1 & 0.8 & 0.9 & 2.1 & 1.6 & 1.4 & 0.6 & 0.0 & 1.0 & 65 & 0.7 \\
1516 & 0.3 & 0.2 & 0.6 & 1.9 & 0.2 & 0.2 & 0.6 & 0.3 & 1.5 & 0.3 & 0.4 & 0.2 & 1.7 & 0.0 & 0.2 & 1.0 & 1.4 & 68 & 0.2 \\
1521 & 1.5 & \nodata & 0.5 & 2.6 & \nodata & 0.4 & 1.0 & 0.1 & 1.7 & 0.5 & 0.4 & 0.7 & 0.0 & 0.0 & 0.1 & 0.2 & 1.6 & 93 & 0.5 \\
1568 & 0.6 & 0.8 & 1.5 & 8.1 & 1.9 & 4.1 & 4.2 & 0.8 & 0.6 & 0.2 & 0.4 & 0.5 & 0.7 & 0.7 & 0.3 & 1.2 & 2.7 & 265 & 5.4 \\
1595 & 0.3 & \nodata & 0.7 & 2.4 & \nodata & 0.5 & 0.3 & 0.7 & 0.8 & 0.2 & 0.3 & 1.0 & 0.0 & 4.2 & 0.1 & 1.6 & 1.8 & 60 & 0.2 \\
1608 & 2.9 & 0.2 & 1.3 & 5.2 & 1.2 & 1.3 & 0.6 & 0.1 & 0.6 & 0.0 & 0.2 & 0.1 & 1.3 & 0.0 & 0.2 & 0.0 & 1.2 & 92 & 0.6 \\
\enddata
\renewcommand{\baselinestretch}{1.2}
\tablenotetext{a}{Interstellar absorption H column density, in units of 
$10^{21}$ cm$^{-2}$.
}
\tablenotetext{b}{Volume emission measure for each thermal component, divided 
by $4 \pi D^2$ (with $D$ the source distance), in units of $10^{-10}$ cm$^{-5}$.
}
\tablenotetext{c}{Absorbed source X-ray flux, in the 2 -- 8 keV (hard) band,
in units of $10^{-13}$ erg cm$^{-2}$ s$^{-1}$.
}
\end{deluxetable}

%% file: tab3.tex
\begin{deluxetable}{crcccccc}
\tablecaption{Abundances in ONC X-ray bright stars
\label{tab:abres}}
\tablewidth{0pt}
\tablehead{
        &     & \multicolumn{3}{c}{Low-absorption sample} &
\multicolumn{3}{c}{Count-limited sample} \\
        & FIP & median & \multicolumn{2}{c}{68\% range} & median &
\multicolumn{2}{c}{68\% range} \\
Element & (eV) & \multicolumn{6}{c}{(\citet{ag89} solar abundance units)}
}
\startdata
Ca & 6.11 & 0.00 & 0.00 & 0.90 & 0.00 & 0.00 & 0.54 \\
Ni & 7.64 & 1.32 & 0.31 & 2.51 & 1.37 & 0.65 & 2.20 \\
Mg & 7.65 & 0.25 & 0.03 & 0.58 & 0.26 & 0.09 & 0.49 \\
Fe & 7.90 & 0.22 & 0.12 & 0.37 & 0.22 & 0.14 & 0.33 \\
Si & 8.15 & 0.33 & 0.14 & 0.70 & 0.30 & 0.13 & 0.55 \\
S & 10.36 & 0.75 & 0.43 & 1.23 & 0.74 & 0.45 & 1.13 \\
O & 13.62 & 0.57 & 0.25 & 1.26 & 0.60 & 0.29 & 1.00 \\
Ar & 15.76 & 0.76 & 0.20 & 1.52 & 0.76 & 0.49 & 1.33 \\
Ne & 21.56 & 1.34 & 0.59 & 2.23 & 1.14 & 0.65 & 2.27 \\
\enddata
\end{deluxetable}

%% file: ms_final.bbl
\begin{thebibliography}{}

\bibitem[{{Anders} \& {Grevesse}(1989)}]{ag89}
{Anders}, E., \& {Grevesse}, N. 1989, Geochimica et Cosmochimica Acta,
53, 197
\bibitem[Antia \& Basu(2005)]{ab05}
Antia, H. M., \& Basu, S. 2005, \apj, 620, L129
\bibitem[Arge \& Mullan(1998)]{am98}
Arge, C. N., \& Mullan, D. J. 1998, Sol. Phys., 182, 293
\bibitem[{{Argiroffi} {et~al.}(2004){Argiroffi}, {Drake}, {Maggio},
{Peres}, {Sciortino}, \& {Harnden}}]{a+04}
{Argiroffi}, C., {Drake}, J.~J., {Maggio}, A., {Peres}, G., {Sciortino},
S., \& {Harnden}, F.~R. 2004, \apj, 609, 925
\bibitem[{Argiroffi} {et~al.}(2005)]{a+05}
{Argiroffi}, C., {Maggio}, A., {Peres}, G., {Stelzer}, B., \&
{Neuh{\"a}user}, R. 2005, \aap, 439, 1149
\bibitem[Arnaud(1996)]{a96}
Arnaud, K. A. 1996, in ASP Conf. Ser. 101, Data Analysis Software and
Systems V, ed. G. H. Jacoby \& J. Barnes (San Francisco: ASP), 17
\bibitem[Asplund(2005)]{a05}
Asplund, M. 2005, \araa, 43, 481
\bibitem[Asplund et al.(2005)Asplund, Grevesse, \& Sauval]{ags05}
Asplund, M., Grevesse, N., Sauval, A. J. 2005, in ASP Conf. Ser. 336,
Cosmic Abundances as Records of Stellar Evolution and Nucleosynthesis,
ed. T. G. Barnes III and F. N. Bash, 25
\bibitem[Audard et al.(2001)Audard, G\"udel, \& Mewe]{agm01}
Audard, M., G\"udel, M., Mewe, R. 2001, \aap, 365, L318
\bibitem[Audard et al.(2003)]{ags+03}
Audard, M., G{\"u}del, M., Sres, A., Raassen, A. J. J., \& Mewe, R.
2003, \aap, 398, 1137
\bibitem[Bahcall et al.(2005)Bahcall, Basu, \& Serenelli]{bbs05}
Bahcall, J. N., Basu, S., Serenelli, A. M. 2005, \apj 631, 1281
\bibitem[Brinkman et al.(2001)]{b+01}
Brinkman, A. C.~et al.\ 2001, \aap, 365, L324 
\bibitem[Cunha et al.(2006)Cunha, Hubeny, \& Lanz]{chl06}
Cunha, K., Hubeny, I., \& Lanz, T. 2006, \apj, 647, L143
\bibitem[Cunha \& Lambert(1992)]{cl92}
Cunha, K., \& Lambert, D. L. 1992, \apj, 399, 586
\bibitem[Cunha \& Lambert(1994)]{cl94}
Cunha, K., \& Lambert, D. L. 1994, \apj, 426, 170
\bibitem[Cunha \& Smith(2005)]{cs05}
Cunha, K., \& Smith, V. V. 2005, \apj, 626, 425
\bibitem[Cunha et al.(1998)Cunha, Smith, \& Lambert]{csl98}
Cunha, K., Smith, V. V., \& Lambert, D. L. 1998, \apj, 493, 195
\bibitem[Drake et al.(2001)]{dbk+01}
Drake, J. J., Brickhouse, N. S., Kashyap, V., Laming, J.
M., Huenemoerder, D. P., Smith, R., Wargelin, B. J.
2001, \apj, 548, 81
\bibitem[Drake \& Testa(2005)]{dt05}
Drake, J. J., Testa, P. 2005, \nat, 436, 525
\bibitem[Drake et al.(2005)Drake, Testa, \&Hartmann(2005)]{dth05}
Drake, J. J., Testa, P., Hartmann, L. 2005, \apj, 627, 149
\bibitem[Esteban et al.(2004)]{e+05}
Esteban, C., Peimbert, M., Garc\'ia-Rojas, J., Ruiz, M. T., Peimbert,
A.,
Rodr\'iguez, M. 2004, \mnras, 355, 229
\bibitem[Favata et al.(2005)]{ffr+05}
Favata, F., Flaccomio, E., Reale, F., Micela, G., Sciortino, S., Shang,
H., Stassun, K. G., Feigelson, E. D. 2005, \apj, 160, 469
\bibitem[Favata \& Micela(2003)]{fm03}
Favata, F., \& Micela, G. 2003, SSRv, 108, 577
\bibitem[Feigelson et al.(2005)]{fgt+05}
Feigelson, E. D., Getman, K., Townsley, L., Garmire,
G., Preibisch, T., Grosso, N., Montmerle, T., Muench,
A., McCaughrean, M. 2005, \apjs, 160, 379
\bibitem[Feldman \& Laming(2000)]{fl00}
Feldman, U., \& Laming, J. M. 2000, \physscr, 61, 222
\bibitem[Garcia-Alvarez et al.(2006)]{gdb+06}
García-Alvarez, D., Drake, J. J., Ball, B., Lin, L., Kashyap,
V. L. 2006, \apj, 638, 1028
\bibitem[Garmire et al.(2003)]{g+03}
Garmire, G. P., Bautz, M. W., Ford, P. G., Nousek, J. A.,
Ricker, G. R., Jr. 2003, SPIE, 4851, 28
\bibitem[{Getman} {et~al.}(2005)]{g+05}
{Getman}, K.~V., {Feigelson}, E.~D., {Grosso}, N., {McCaughrean}, M.~J.,
  {Micela}, G., {Broos}, P., {Garmire}, G., \& {Townsley}, L. 2005,
\apjs, 160, 319
\bibitem[G\"udel(2004)]{g04}
G\"udel, M. 2004, \aapr, 12, 71
\bibitem[Hillenbrand(1997)]{h97}
Hillenbrand, L. A. 1997, \aj, 113, 1733
\bibitem[Kastner et al.(2002)]{khs+02}
Kastner, J. H., Huenemoerder, D. P., Schulz, N. S., Canizares,
\bibitem[Kastner et al.(2005)]{kfg+05}
Kastner, J. H., Franz, G., Grosso, N., Bally, J.,
McCaughrean, M. J., Getman, K., Feigelson, E. D., Schulz,
N. S. 2005, \apjs, 160, 511
C. R., Weintraub, D. A. 2002, \apj, 567, 434
\bibitem[Laming(2004)]{l04}
Laming, J. M. 2004, \apj, 614, 1063
\bibitem[Lampton et al.(1976)Lampton, Margon, \& Bowyer]{lmb76}
Lampton, M., Margon, B., \& Bowyer, S. 1976, \apj, 208, 177
\bibitem[Kenyon \& Hartmann(1995)]{kh95}
Kenyon, S. J., Hartmann, L. 1995, \apjs, 101, 117
\bibitem[Maggio et al.(2005)]{m+05}
Maggio, A., Drake, J. J., Favata, F., G\"udel, M.
2005, in ESA SP 560, Proceedings of
  the 13th Cambridge Workshop on Cool Stars, Stellar Systems and the
Sun, ed. F.~Favata, G.A.J. Hussain, B. Battrick, 129
\bibitem[Mewe et al.(1995)Mewe, Kaastra, Liedahl]{mkl95}
Mewe, R., Kaastra, J. S., Liedahl, D. A. 1995, Legacy, 6, 16
\bibitem[{{Meyer} {et~al.}(1997){Meyer}, {Calvet}, \&
{Hillenbrand}}]{mch97}
{Meyer}, M.~R., {Calvet}, N., \& {Hillenbrand}, L.~A. 1997, \aj, 114,
288
\bibitem[Morrison \& McCammon(1983)]{mm83}
Morrison, R., McCammon, D. 1983, \apj, 270, 119
\bibitem[Nordon et al.(2006)Nordon, Behar, \& G\"udel]{nbg06}
Nordon, R., Behar, E., G\"udel, M. 2006, \aap, 446, 621
\bibitem[Osten et al.(2004)]{oba+04}
Osten, Rachel A.,g Brown, Alexander; Ayres, Thomas R.; Drake,
Stephen A.,g Franciosini, Elena; Pallavicini, Roberto; Tagliaferri,
Gianpiero,g Stewart, Ron T.; Skinner, Stephen L.; Linsky, Jeffrey L.
2004, \apjs, 153, 317
\bibitem[Sanz-Forcada et~al.(2004)Sanz-Forcada, Favata, \& Micela]{sfm04}
{Sanz-Forcada}, J., {Favata}, F., \& {Micela}, G. 2004, \aap, 416, 281
\bibitem[Santos et al.(2005)]{sim+05}
Santos, N. C., Israelian, G., Mayor, M., Bento, J. P., Almeida, P. C.,
Sousa, S. G., Ecuvillon, A. 2005, \aap, 437, 1127
\bibitem[Sanz Forcada et al.(2003a)Sanz Forcada, Brickhouse, \&
Dupree]{sbd03}
Sanz Forcada, J., Brickhouse, N. S., \& Dupree, A. K. 2003, \apjs, 145,
147
\bibitem[{{Sanz-Forcada} {et~al.}(2003b){Sanz-Forcada}, {Maggio}, \&
  {Micela}}]{smm03}
{Sanz-Forcada}, J., {Maggio}, A., \& {Micela}, G. 2003, \aap, 408, 1087
\bibitem[Schmitt \& Ness(2004)]{sn04}
Schmitt, J. H. M. M., \& Ness, J.-U. 2004, \aap, 415, 1099
\bibitem[Schmitt et al.(2005)]{srn+05}
Schmitt, J. H. M. M., Robrade, J., Ness, J.-U., Favata, F., Stelzer, B.
2005, \aap, 432, 35
\bibitem[{{Siess} {et~al.}(2000){Siess}, {Dufour}, \&
  {Forestini}}]{sdf00}
{Siess}, L., {Dufour}, E., \& {Forestini}, M. 2000, \aap, 358, 593
\bibitem[Sim\'on-D\'iaz et al.(2006)]{sd+06}
Sim\'on-D\'iaz, S., Herrero, A., Esteban, C., Najarro, F. 2006, \aap,
448, 351
\bibitem[Smith et al.(2001)]{sbl+01}
Smith, R. K., Brickhouse, N. S., Liedahl, D. A., \& Raymond, J. C.
2001, \apjl, 556, 91
\bibitem[Stassun et al.(2006)]{svff06}
Stassun K. G., van den Berg, M., Feigelson, E., Flaccomio, E. 2006,
\apj, 649, 914
\bibitem[Stelzer \& Schmitt(2004)]{ss04}
Stelzer, B., Schmitt, J. H. M. M. (2004), \aap, 418, 687
\bibitem[Stelzer et al.(2007)]{sfb+06}
Stelzer, B., Flaccomio, E., Briggs, K., Micela, G., Scelsi, L., Audard,
M., Pillitteri, I., G\"udel, M. 2007, \aap, in press
\bibitem[Schwadron et al.(1999)Schwadron, Fisk, \& Zurbuchen]{s+99}
Schwadron, N. A., Fisk, L. A., \& Zurbuchen, T. H. 1999, \apj, 521, 859
\bibitem[Weisskopf et al.(2002)]{w+02}
Weisskopf, M. C., Brinkman, B., Canizares, C., Garmire, G., Murray, S.,
Van Speybroeck, L. P. 2002, \pasp, 114, 1
\bibitem[Wolk et al.(2005)]{whf+05}
Wolk, S. J., Harnden, F. R., Jr., Flaccomio, E., Micela, G.,
Favata, F., Shang, H., Feigelson, E. D. 2005, \apjs, 160, 423
\bibitem[Wood \& Linsky(2006)]{wl06}
Wood, B. E., Linsky, J. L. 2006, \apj, 643, 444

\end{thebibliography}
